\RequirePackage{fix-cm}
\documentclass[twocolumn,epjc3]{svjour3}
\RequirePackage{flushend}
\RequirePackage[colorlinks,citecolor=blue,urlcolor=blue,linkcolor=blue]{hyperref}
\RequirePackage{cite}
\RequirePackage{graphicx}
\RequirePackage{amssymb}
\RequirePackage{amsmath}
\begin{document}

\title{
Fermion-soliton scattering in a modified $\mathbb{CP}^{1}$ model
}
\author{A.Yu.~Loginov\thanksref{addr1,e1}}
\thankstext{e1}{e-mail: a.yu.loginov@tusur.ru}
\institute{Laboratory of Applied Mathematics and Theoretical Physics, Tomsk State
           University of Control Systems and Radioelectronics, 634050 Tomsk, Russia \label{addr1}}
\date{Received: date / Accepted: date}
\maketitle

\begin{abstract}
The scattering of fermions  in  the  background  field of a topological soliton
of the modified  $(2 + 1)$-dimensional $\mathbb{CP}^{1}$  model is studied here
both analytically and numerically.
Unlike the  original  $\mathbb{CP}^{1}$  model,  the Lagrangian of the modified
model contains a potential term.
Due to this, a  dilatation   zero  mode  of the topological soliton disappears,
which results in stability of the fermion-soliton system.
The symmetry properties  of  the  fermion-soliton  system  are established, and
the asymptotic forms of fermionic radial wave functions are studied.
Questions related to the bound states of the  fermion-soliton  system  are then
discussed.
General  formulae  describing  the   scattering   of  fermions  are  presented.
The amplitudes of the  fermion-soliton scattering are obtained in an analytical
form within  the  framework  of  the  Born  approximation,  and  their symmetry
properties and asymptotic  forms  are studied.
The energy levels of the fermionic bound states and the partial phase shifts of
fermionic scattering   are    obtained    by    numerical    methods,   and the
ultrarelativistic limits of the  partial  phase  shifts are found.

\end{abstract}

\section{Introduction} \label{sec:I}

Spatially localised nondissipative solutions  with finite energy and nontrivial
topology exist   in   many  models  of  field  theory  \cite{Manton,  Weinberg,
  Zakrzewski}.
Known as topological solitons, these solutions play  an important role in field
theory,  high-energy   physics,   condensed   matter   physics,  cosmology, and
hydrodynamics.
Among  these,  we  can  distinguish   a  class  of  planar topological solitons
of $(2 + 1)$-dimensional field models.
The vortices of the effective  theory of superconductivity \cite{abr_jetp_1957}
and  of     the     $\left(2 + 1\right)$-dimensional    Abelian    Higgs  model
\cite{nielsen_npb_1973}  are   probably   the  best  known  planar  topological
solitons.
The topological soliton  of the $(2+1)$-dimensional nonlinear $O\left(3\right)$
sigma model, called  a  lump,   is  another   well-known   example  of a planar
topological soliton \cite{bpl_jetpl_1975}.

Nonlinear sigma models can  also  be  formulated  for  orthogonal groups $O(N)$
with $N \ge 4$,  but  unlike  the  $O(3)$  sigma  model,  these  models have no
topological solitons.
There is also another family  of  nonlinear  field  models whose properties are
similar  to  those  of  the  nonlinear  $O(N)$  sigma  models in many respects;
these are  the  so-called $\mathbb{CP}^{N - 1}$  models \cite{cremmer_plb_1978,
 eich_npb_1978, glpr_lmp_1978, glpr_plb_1978}.
For $N = 2$, the $\mathbb{CP}^{1}$  model  is  equivalent  to  the $O(3)$ sigma
model, whereas for $N  \ge  3$,  the  $\mathbb{CP}^{N - 1}$  model  is a better
generalisation of the $O(3)$ sigma model than the $O(N+1)$  sigma model, as  it
continues  to   have   topological   soliton   solutions  \cite{dadda_npb_1978,
 witten_npb_1979}.

Soon after  their  appearance   in   the   late  1970s,  it  was  realised that
two-dimensional   $\mathbb{CP}^{N - 1}$   models   could   be   used   to study
nonperturbative effects  in  four-dimensional  Yang-Mills models, as these  two
types of models have many properties in common, such as conformal invariance at
the classical  level,   asymptotic    freedom    in    the   ultraviolet region
\cite{polyakov_plb_1975},  and   strong   coupling   in   the  infrared region.
Furthermore, a topological term and  instanton  solutions \cite{dadda_npb_1978,
 witten_npb_1979} exist  for  two  types  of   models,  resulting  in a complex
structure of the vacuum at the quantum level.
It is clear that the lower dimensionality of the  $\mathbb{CP}^{N - 1}$  models
simplifies the analysis  of  nonperturbative  effects  in  the strong  coupling
regime  compared  to  the  more  complex   four-dimensional  Yang-Mills models.

Two-dimensional $\mathbb{CP}^{N-1}$ models  can  be used to describe low-energy
dynamics on the world  sheet  of  non-Abelian  vortex  strings  in  a class  of
four-dimensional    gauge   theories   \cite{hantng_jhep_2003,  auzzi_npb_2003,
shifyung_prd_2004,  hantng_jhep_2004,  shifyung_rmp_2007,  tong_anp_2009}.
In addition,  $\mathbb{CP}^{N - 1}$  models  have  interesting  applications in
various fields of condensed matter physics \cite{Tsvelik}, and  particularly in
relation to ferromagnetism, the Hall effect, and the Kondo effect.
They also find  application  in  the  study  of  the  fermion  number violation
realised via a sphaleron transition at high temperature  \cite{mwipf_prd_1989}.

The static energy functional of the $(2+1)$-dimensional $\mathbb{CP}^{1}$ model
is invariant under scale transformations, meaning that the soliton solutions of
the model (lumps) depend on an  arbitrary  scale  parameter that determines the
soliton size.
At the same time, the energy of  a  lump  does  not depend on its spatial size.
As a result,  in  addition  to  two  translational  zero   modes, the lump also
possesses a dilatation zero mode.
This is a source  of   instability  of  the  lump  in  dynamic processes, since
collisions between  lumps  or   interactions between   the lump and bosonic and
fermionic fields can lead  to  the radius of the lump tending either to zero or
to      infinity      \cite{leese_prd_1989,     lpz_nln_1990,    pzkr_nln_1996,
 ioannidou_nln_1997}.

There are several ways in which the $\mathbb{CP}^{1}$  model can be modified to
remove the size instability of a lump.
One of them, which was proposed in  Ref.~\cite{leese_npb_1991}, involves adding
a potential  term  of  a  certain   type   to   the  Lagrangian of the original
$\mathbb{CP}^{1}$ model.
This breaks  the  scale  invariance  of  the  original  model,  leading  to the
disappearance of the dilatation zero mode.
At the same time, it  follows  from  Derrick's  theorem \cite{derrick_jmp_1964}
that there can no longer be  static  lump  solutions,  since the potential term
will cause them to collapse.
However, it was shown  in  Ref.~\cite{leese_npb_1991}  that time-dependent lump
solutions can be constructed in this case.
These solutions,  called  Q-lumps  due   to  their  similarities  with  Q-balls
\cite{coleman_npb_1985}, have  the  same  form  as  the  original  lumps of the
$\mathbb{CP}^{1}$ model, except  that  their  phase changes linearly with time.
Due to this, Q-lumps carry a conserved Noether charge, which prevents them from
collapsing.

The $\mathbb{CP}^{N-1}$ models can be extended to  include  fer- mionic fields,
either by  a  supersymmetric   extension  of  the  $\mathbb{CP}^{N - 1}$  model
\cite{cremmer_plb_1978, witten_npb_1979, dadda_npb_1979} or by minimal coupling
between fermionic  fields  and  a  composite gauge field of the $\mathbb{CP}^{N
- 1}$ model \cite{abdalla_prd_1982}.
The supersymmetric extension of the $\mathbb{CP}^{N-1}$ model involves Majorana
fermionic fields that satisfy nontrivial constraints, whereas the minimal model
deals with unconstrained Dirac fermionic fields.
In the present paper, we study, within the  background  field  approximation, a
fermion-soliton system  of  the  $\mathbb{CP}^{1}$ model with a potential term.
In this model, Dirac fermionic fields interact minimally with a composite gauge
field.
The results obtained here can  be  used  to  describe  the  interaction between
fermions  and  two-dimensional  or  thread-like  three-dimensional  topological
defects in condensed matter physics.

This  paper  is  structured as follows.
In Sec.~\ref{sec:II}, we briefly  describe  the  Lagrangian,  symmetries, field
equations,  and  topological solitons  of the modified $\mathbb{CP}^{1}$ model.
In Sec.~\ref{sec:III},  we   explore   fermion-soliton  scattering  within the
background field approximation.
We establish the symmetry  properties of  the fermion-soliton system, and study
the asymptotic forms of fermionic radial wave functions.
We also consider some questions  concerning fermionic bound states, and present
general formulae for fermion-soliton scattering.
In Sec.~\ref{sec:IV},   we   present   an   analytical  description  of fermion
scattering within the framework of the Born approximation.
In Sec.~\ref{sec:V}, we present  numerical  results,  based  on which we obtain
several analytical expressions related to fermion-soliton scattering.
In particular, we find expressions for  the  fermionic  partial phase shifts in
the ultrarelativistic limit.
In the final section, we briefly  summarise the results obtained in the present
work.
In Appendix A, we derive expressions for  the fermionic partial phase shifts in
the semiclassical approximation.

Throughout the paper, the natural units $c = 1$ and $\hbar = 1$ are used.

\section{Lagrangian,  field   equations  and topological solitons of the model}
                                                                 \label{sec:II}
The Lagrangian density of the model considered here has the form
\begin{eqnarray}
\mathcal{L} &=&g^{-1}\left( D_{\mu }n_{a}\right) ^{\ast }D^{\mu
}n_{a}-g^{-1}U\left( \left\vert n_{1}\right\vert,\left\vert n_{2}\right\vert
\right)                                                             \nonumber
  \\
&&+i\bar{\psi}_{a}\gamma^{\mu}D_{\mu }\psi_{a} - M\bar{\psi}
_{a}\psi _{a},                                                     \label{II:1}
\end{eqnarray}
where $g$ is a  coupling  constant, $n_{a}$  is  a  complex  scalar isodoublet,
$\psi_{a}$ is the Dirac fermionic  isodoublet,  and  $M$ is the fermionic mass.
In Eq.~(\ref{II:1}), the  complex  scalar  isodoublet  is  under the constraint
$n_{a}^{\ast }n_{a} = 1$, the quartic potential term
\begin{eqnarray}
U\left(\left\vert n_{1}\right\vert,\left\vert n_{2}\right\vert \right)
& = & 2^{-2}\alpha ^{2}\left[ 1-\bigl( \left\vert n_{1}\right\vert
^{2}-\left\vert n_{2}\right\vert ^{2}\bigr) ^{2}\right]             \nonumber
 \\
& = & \alpha ^{2}\left\vert n_{1}\right\vert ^{2}\left\vert
n_{2}\right\vert ^{2},                                             \label{II:2}
\end{eqnarray}
where $\alpha$ is a parameter  with  the  dimension  of mass, and the covariant
derivatives of the fields are defined as
\begin{subequations}                                               \label{II:3}
\begin{eqnarray}
D_{\mu}n_{a} &=&\partial_{\mu}n_{a} + i A_{\mu}n_{a}              \label{II:3a}
 \\
\text{and} \phantom{oooo} & &                                       \nonumber
 \\
D_{\mu}\psi_{a} &=&\partial_{\mu}\psi_{a} + i A_{\mu}\psi_{a},    \label{II:3b}
\end{eqnarray}
\end{subequations}
where $A_{\mu}$ is a vector gauge field.
The Lagrangian density  in  Eq.~(\ref{II:1})  describes  the  $\mathbb{CP}^{1}$
model that possesses the quartic potential $U\left(\left\vert n_{1}\right\vert,
\left\vert n_{2}\right\vert \right)$ and interacts minimally with the fermionic
isodoublet $\psi_{a}$.

The field equations for model (\ref{II:1})  are  obtained by varying the action
$S = \int \mathcal{L}d^{2}xdt$  in  the  fields  $n_{a}$, $\bar{\psi}_{a}$, and
$A_{\mu}$, and taking into  account  the  constraint $n_{a}^{\ast}n_{a} = 1$ by
means of the Lagrange multiplier method:
\begin{eqnarray}
D_{\mu}D^{\mu}n_{a}-\left(n_{b}^{\ast} D_{\mu}D^{\mu}n_{b}\right) n_{a}+
\mathcal{P}_{ab} \partial_{n_{b}^{\ast }}U &=&0,                  \label{II:4a}
 \\
\left( i\gamma ^{\mu }D_{\mu }-M\right) \psi _{a} &=&0,           \label{II:4b}
 \\
A_{\mu }-in_{a}^{\ast }\partial _{\mu }n_{a}-\frac{g}{2}\bar{\psi}
_{a}\gamma _{\mu }\psi _{a} &=&0,                                 \label{II:4c}
\end{eqnarray}
where the  projector $\mathcal{P}_{ab} = \mathbb{I}_{ab} - n_{a}n_{b}^{\ast }$.
Eq.~(\ref{II:4c}) tells us that that the gauge  field $A_{\mu}$ is expressed in
terms of the fields $n_{a}$ and $\psi_{a}$, meaning  that it is not dynamic but
auxiliary.

In the absence of the potential  term, the Lagrangian (\ref{II:1}) is invariant
under transformations of  a  $SU(2)\times U(1)$ group, where the first (second)
factor corresponds to global (local) transformations.
The presence of the potential  term  leads  to   breaking of the $SU(2)$ global
factor to a $U(1)$  subgroup corresponding to the generator $t_{3}=\tau_{3}/2$,
whereas the $U(1)$ gauge factor remains unbroken.
The Noether currents  that  correspond  to  the first and second factors of the
symmetry group $U(1)\times U(1)$ are
\begin{eqnarray}
j_{3}^{\mu } & = & - i g^{-1}\text{Tr}\left[ \tau_{3}n\overleftrightarrow{D}
^{\mu}n^{\ast}\right]+\text{Tr}\left[\tau_{3}\bar{\psi }\gamma
_{\mu }\psi \right]                                               \label{II:5a}
  \\
\text{and} \phantom{oooo} & &                                       \nonumber
  \\
j^{\mu } & = &-ig^{-1}\text{Tr}\left[ n\overleftrightarrow{D}^{\mu
}n^{\ast }\right]
+ \text{Tr}\left[ \bar{\psi }\gamma _{\mu }\psi \right],          \label{II:5b}
\end{eqnarray}
respectively.
In Eqs.~(\ref{II:5a}) and (\ref{II:5b}),  the  trace is over the indices of the
isodoublets $n$ and $\psi$ and the Pauli matrix $\tau_{3}$.

Using the well-known formula  $T_{\mu \nu }\!=\!2 \partial \mathcal{L}/\partial
\eta^{\mu \nu }\!  -  \eta^{\mu \nu } \mathcal{L}$,  we  obtain  the  symmetric
energy-momentum tensor for a bosonic field configuration of  model (\ref{II:1})
as
\begin{eqnarray}
T_{\mu \nu } &=&2g^{-1}\left( D_{\mu }n_{a}\right) ^{\ast }D_{\nu
}n_{a}-\eta _{\mu \nu }g^{-1}                                       \nonumber
 \\
&&\times \left[ \left( D_{\sigma }n_{a}\right) ^{\ast }D^{\sigma
}n_{a}-U\left( \left\vert n_{1}\right\vert ,\left\vert n_{2}\right\vert
\right) \right].                                                   \label{II:6}
\end{eqnarray}
Eq.~(\ref{II:6})  tells  us  that  for  the  energy  $E = \int T_{00}d^{2}x$ of
a field configuration  to  be  finite,  the  potential $U$ must tend to zero at
spatial infinity.
It then follows from  Eq.~(\ref{II:2}) that at spatial infinity, we have either
$n_{1} = e^{i f_{1}\left(\theta \right) },\, n_{2}=0$ or $n_{1} = 0, \, n_{2} =
e^{i f_{2}\left( \theta \right)}$,  where  $\theta$  is   the  polar  angle and
$f_{1,2}(\theta)$  are  periodic  functions of $\theta$.
In any case, each finite energy field  configuration  of model (\ref{II:1}) can
be attributed to one of the classes  of  the  homotopy  group $\pi_{1}(S^{1}) =
\mathbb{Z}$.
Hence, the finite  energy  field  configurations  of  model (\ref{II:1}) can be
labelled by an  integer  $n$, called the winding number, as explicitly given in
Ref.~\cite{dadda_npb_1978}:
\begin{equation}
n = -\frac{1}{2\pi}\int\limits_{S^{1}}A_{i}dx^{i} =
    -\frac{1}{2\pi}\int \epsilon_{ij}\partial_{i}A_{j} d^{2}x,     \label{II:7}
\end{equation}
where $A_{i} = i n_{a}^{\ast} \partial_{i} n_{a}$ and  $\epsilon_{i j}$  is the
two-dimensional antisymmetric tensor with $\epsilon_{1 2} = 1$.

It follows from Derrick's theorem \cite{derrick_jmp_1964} that  the presence of
the potential term excludes the  existence  of  static soliton solutions in the
$(2+1)$-dimensional model (\ref{II:1}).
At the same time, the presence  of the potential term does not prohibit soliton
solutions with nontrivial time dependence.
Indeed, it can be shown  that  a  rather  specific  form  of  the  potential in
Eq.~(\ref{II:2}) allows us to rewrite the obvious inequality
\begin{eqnarray}
&&\int \left[\left( D_{i}n_{a}\pm i\epsilon
_{ij}D_{j}n_{a}\right) ^{\ast }\left( D_{i}n_{a}\pm i\epsilon
_{ij}D_{j}n_{a}\right)/2 +\right.                                  \label{II:8}
  \\
&& \left. \left(\partial_{t}n_{a}+i\alpha \left(t_{3}\right)
_{ab}n_{b}\right) ^{\ast }\left( \partial _{t}n_{a} + i \alpha \left(
t_{3}\right)_{ab}n_{b}\right) \right] d^{2}x \geq 0                 \nonumber
\end{eqnarray}
in the form
\begin{equation}
E - 2 \pi g^{-1} \left\vert n\right\vert -\left\vert \alpha \right\vert
\left\vert Q_{3}\right\vert/2 \geq 0,                              \label{II:9}
\end{equation}
where $E = \int T_{00}d^{2}x$, $n = -(2\pi)^{-1} \int \epsilon_{ij}\partial_{i}
A_{j}  d^{2}x$,  and  $Q_{3}  =   - i  g^{-1}  \int  \text{Tr} \bigl[\tau_{3} n
\overleftrightarrow{D}^{\mu} n^{\ast}\bigr] d^{2}x$ are the energy, the winding
number, and the Noether charge of the field configuration, respectively.
It follows from Eq.~(\ref{II:8}) that the saturation of inequality (\ref{II:9})
is only possible for field  configurations that satisfy the Bogomolny equations
\begin{eqnarray}
\partial_{t}n_{a} + i \alpha \left(t_{3}\right)_{ab}n_{b} &=&0,  \label{II:10a}
  \\
D_{i}n_{a}\pm i\epsilon _{ij}D_{j}n_{a} &=&0.                    \label{II:10b}
\end{eqnarray}
It can be shown that  solutions  to Eqs.~(\ref{II:10a}) and (\ref{II:10b}) also
satisfy field equation (\ref{II:4a})  if  we  neglect  the  contribution of the
fermionic fields in Eq.~(\ref{II:4c}).

Eq.~(\ref{II:10a}) determines  the  time  dependence  $n_{a} \left(t,\mathbf{x}
\right) = \exp\left[-i\alpha t t_{3}\right]_{a b} n_{b}\left(\mathbf{x} \right)
\phantom{i}$ of the  field  configurations,  and  Eq.~(\ref{II:10b}) determines
their spatial dependence.
All solutions  to  Eq.~(\ref{II:10b})  are   well  known  \cite{dadda_npb_1978,
 witten_npb_1979} and  can  be obtained in analytical form.
In   particular, the  $\mathbb{Z}_{\left\vert n \right\vert}$ symmetric soliton
solution with winding number $n$ can be written as
\begin{equation}
n_{a}\left( t,\rho ,\theta \right) =\exp \left[ -i\alpha t t_{3} \right]_{ab}
\frac{\lambda ^{\left\vert n\right\vert }u_{b}+\rho ^{\left\vert
n\right\vert }e^{in\theta }v_{b}}{\left( \lambda ^{2\left\vert n\right\vert
}+\rho ^{2\left\vert n\right\vert }\right) ^{1/2}},               \label{II:11}
\end{equation}
where $\rho$  and  $\theta$  are  polar coordinates, $u^{T} = \left(1,0\right)$
and  $v^{T}=\left( 0, 1\right)$ are orthonormal isospinors, and  $\lambda$ is a
parameter that determines the effective size of the soliton.

Having obtained  the   analytical  form  in  Eq.~(\ref{II:11}),  we  can derive
analytical expressions for the auxiliary gauge field
\begin{equation}
A_{\mu }=n\frac{\rho ^{2\left( \left\vert n\right\vert -1\right) }}{\rho
^{2\left\vert n\right\vert }+\lambda ^{2\left\vert n\right\vert }}\left[
\frac{\alpha }{2n}\frac{\lambda ^{2\left\vert n\right\vert }-\rho
^{2\left\vert n\right\vert }}{\rho ^{2\left( \left\vert n\right\vert
-1\right) }},y,-x\right],                                         \label{II:12}
\end{equation}
the Noether charge
\begin{equation}
Q_{3}=\int j_{3}^{0}d^{2}x=\frac{4\pi ^{2}\alpha \lambda ^{2}}{gn^{2}\sin
\left( \pi /\left\vert n\right\vert \right)},                     \label{II:14}
\end{equation}
and the energy 
\begin{equation}
E = \int T_{00}d^{2}x = 2 \pi \left\vert n \right\vert g^{-1}
 + \alpha Q_{3}/2                                                 \label{II:15}
\end{equation}
of the  soliton  solution,  where  we  factor  out  the  common  factor  of the
components of the auxiliary gauge field $A_{\mu}$ for brevity.
Eqs.~(\ref{II:14}) and (\ref{II:15})  tell us that $Q_{3}$ and $E$ are infinite
when the winding number $n = \pm 1$.
Hence, there are no soliton solutions for $\left \vert n \right \vert = 1$, and
$\left \vert n \right \vert= 2$ is the minimum possible value for the magnitude
of the winding number.
We note that $A_{0}$ does not depend  on  the  sign  of  $n$ but on the sign of
$\alpha$, and  that  $\text{sign}(Q_{3}) = \text{sign}(\alpha)$,  which implies
that $\alpha Q_{3} > 0$.
We also note that in  Eq.~(\ref{II:15}),  the  topological  part  of the energy
increases  linearly   with an increase in  $\left\vert n \right\vert$,  whereas
the  Noether (kinetic) part of the energy decreases  monotonically and tends to
zero $\propto \left \vert n \right \vert^{-1}$  as  $\left \vert n \right \vert
\rightarrow \infty$.

From Eqs.~(\ref{II:14})  and  (\ref{II:15}),   we   find   that  the derivative
\begin{equation}
dE/dQ_{3} = \alpha.                                              \label{II:15b}
\end{equation}
We see that the  derivative  of  the  energy of the soliton with respect to its
Noether charge is proportional to  the  phase  frequency  of  rotation  of  the
scalar isodoublet.
This behavior is typical for all nontopological solitons, and in particular for
the Q-balls \cite{coleman_npb_1985}.
In view of this similarity, the  soliton  solutions  of  model (\ref{II:1}) are
called Q-lumps \cite{leese_npb_1991}.

It follows from Eq.~(\ref{II:15}) that in the absence of the potential ($\alpha
= 0$), the  energy  $E$  does  not  depend  on  the  parameter  $\lambda$  that
determines the effective size of the soliton.
In this case, the parameter $\lambda$  and  soliton size are arbitrary, and the
soliton solution has a dilatation zero mode.
This situation changes in the presence of a potential ($\alpha \ne 0$); in this
case, the Noether  charge $Q_{3}$ and the   energy  $E$ depend quadratically on
$\lambda$.
Since both $Q_{3}$ and $E$  are  conserved,  the  parameter  $\lambda$  and the
soliton size are fixed, and there is no dilatation zero mode in this case.

Eq.~(\ref{II:11}) tells us that the soliton solution depends on the polar angle
$\theta$ only via the combination $\phi = n \theta + \alpha t/2$.
We see that a point of constant  $\phi$  rotates  with angular velocity $\omega
= -\alpha/\left(2 n\right)$.
It follows that the  soliton  solution  in  Eq.~(\ref{II:11}) possesses angular
momentum
\begin{equation}
J = \int J^{012} d^{2}x = \frac{4\pi^{2}\lambda^{2}\omega}
{g\sin\left(\pi/\left\vert n\right\vert\right)}=-\frac{n}{2}Q_{3},\label{II:16}
\end{equation}
where the  tensor  $J^{\lambda \mu \nu}  =  x^{\mu} T^{\lambda \nu}  -  x^{\nu}
T^{\lambda \mu }$.
Unlike the  Noether  charge $Q_{3}$,  the angular momentum $J$ depends  on  the
sign of the winding number $n$, and tends to the nonzero limit $-\text{sign}(n)
2\pi g^{-1}\alpha\lambda^{2}$ as $\left\vert n \right\vert \rightarrow \infty$.

\section{Fermion-soliton system in the background field approximation}
                                                                \label{sec:III}
A characteristic property of  model (\ref{II:1})  is  that  the auxiliary gauge
field $A_{\mu} =  i n_{a}^{\ast} \partial_{\mu} n_{a} + 2^{-1} g \bar{\psi}_{a}
\gamma_{\mu} \psi_{a}$ is the  sum  of  quadratic  bosonic and fermionic terms.
Hence, field equation (\ref{II:4a}) contains nonlinear terms that are quadratic
and quartic in the fermionic fields.
These terms describe the fermion backreaction on  the bosonic soliton solution.
Furthermore, the Dirac equation (\ref{II:4b})  also contains  nonlinear (cubic)
terms in the fermionic fields.
A consistent description of these  nonlinear  fermionic  terms is possible only
within the framework of QFT.

Here, we consider the fermion-soliton   system   within   the  background field
approximation, i.e., we neglect the fermionic  term  in $A_{\mu}$ in comparison
with the bosonic term.
In this case, the  auxiliary gauge field $A_{\mu} = in_{a}^{\ast}\partial_{\mu}
n_{a}$, there is no fermion backreaction on the soliton, and the Dirac equation
(\ref{II:4b}) becomes linear in the fermionic field.
Estimation of the bosonic and  fermionic  terms in Eq.~(\ref{II:4c}) shows that
neglecting this term is possible under the condition
\begin{equation}
g \ll a\varrho _{F}^{-1},                                         \label{III:1}
\end{equation}
where $a = \min \left( \left\vert  n \right\vert \lambda ^{-1}, \alpha \right)$
and $\varrho_{F}$ is the  two-dimensional  density  of  fermions in an incident
plane wave.
In addition, we  can  neglect  the  cubic fermionic terms in the Dirac equation
(\ref{II:4b}) compared to the mass term under the condition
\begin{equation}
g \ll M \varrho _{F}^{-1},                                        \label{III:2}
\end{equation}
where $M$ is the fermion mass.

The estimates (\ref{III:1}) and (\ref{III:2}) were obtained from an analysis of
the classical field equations in Eqs.~(\ref{II:4a}) -- (\ref{II:4c}).
From the viewpoint of QFT, however, we are  talking  about  the scattering of a
fermion of mass $M$  on  a  $\mathbb{CP}^{1}$  soliton  of mass $M_{\text{s}} =
2 \pi \left\vert n \right\vert g^{-1} + \alpha Q_{3}/2$.
To enable the recoil of the $\mathbb{CP}^{1}$ soliton to be neglected, the mass
$M_{\text{s}}$  must  be  much  larger  than  the  energy  $\varepsilon$ of the
incident fermion, which leads to the inequality
\begin{equation}
g \ll F\varepsilon^{-1} < F M^{-1},                               \label{III:3}
\end{equation}
where $F=2\pi \left\vert n\right\vert +2\pi ^{2}\alpha^{2}\lambda^{2}n^{-2}\csc
\left( \pi /\left\vert n \right\vert \right)$ does not depend on $g$.

The   conditions   (\ref{III:1}),   (\ref{III:2}),  and  (\ref{III:3})  do  not
contradict each other, and can always be met  if  the  coupling constant $g$ is
sufficiently small.
In this case, the  Dirac  equation  (\ref{II:4b})  is  linear  in the fermionic
fields and can be written in the Hamiltonian form
\begin{equation}
i\partial _{t}\psi _{a}=H\psi _{a},                               \label{III:4}
\end{equation}
where the Hamiltonian
\begin{equation}
H = \mathbb{I}A_{0} + \alpha^{k}\left(-i\partial_{k}
+ A_{k}\right) + \beta M,                                         \label{III:5}
\end{equation}
the Dirac matrices
\begin{equation}
\gamma ^{0}=  \sigma_{3},\,
\gamma ^{1}=-i\sigma_{1},\,
\gamma ^{2}=-i\sigma_{2},                                         \label{III:6}
\end{equation}
the matrices $\alpha^{i} = \gamma^{0}\gamma^{i}$  and $\beta = \gamma^{0}$, and
$\mathbb{I}$ is the identity matrix.
Eqs.~(\ref{III:4}) -- (\ref{III:6})  tell  us  that  the  isodoublet components
$\psi_{1}$ and $\psi_{2}$ of the fermionic  field  interact  in  the  same  way
with the soliton background field.
Hence, we will  not consider the isodoublet index of the fermionic field in the
following.

The Dirac   equation  (\ref{III:4})  possesses   several   discrete symmetries.
Indeed, it can easily be shown that if $\psi(t,\mathbf{x})$  is  a  solution to
this equation, then
\begin{equation}
\psi^{P}\left(t,\mathbf{x}\right)=\sigma_{3}\psi \left(t,
-\mathbf{x}\right),                                              \label{III:7a}
\end{equation}
and
\begin{equation}
\psi^{\Pi_{2}T}\left( t,x,y\right)=\psi^{\ast}\left(
-t,x,-y\right)                                                   \label{III:7b}
\end{equation}
are also solutions to this equation.
Furthermore, the transformed spinor field
\begin{equation}
\psi^{C}\left(t,\mathbf{x}\right) =\sigma_{1}\psi^{\ast}\left(
t,\mathbf{x}\right)                                              \label{III:7c}
\end{equation}
is also a  solution  to  the  Dirac  equation  (\ref{III:4})  provided  that we
substitute $A_{\mu}(\mathbf{x}) \rightarrow A^{C}_{\mu}(\mathbf{x}) = - A_{\mu}
(\mathbf{x})$  in the Hamiltonian (\ref{III:5}).
The solutions in Eqs. (\ref{III:7a}), (\ref{III:7b}),  and  (\ref{III:7c}) are
obtained from the original solution $\psi(t, \mathbf{x})$  by means of the $P$,
$\Pi_{2}T$, and  $C$  transformations,  respectively,  where  in  the  combined
transformation $\Pi_{2}T$, the  symbol  $\Pi_{2}$  denotes reflection about the
$Ox_{1}$ axis.

\subsection{\label{subsec:IIIA}  Fermionic radial wave functions}

In Eq.~(\ref{II:12}), the  time  component $A_{0}$ of the auxiliary gauge field
tends to the  nonzero limit $-\alpha/2$  as $\rho \rightarrow \infty$, which is
inconvenient when studying solutions to the Dirac equation (\ref{III:4}).
We therefore perform the $U(1)$ gauge transformation $A_{\mu}\rightarrow A_{0}-
\partial _{\mu }\Lambda,\, \psi \rightarrow e^{i\Lambda}\psi$, where $\Lambda =
-\alpha t/2$.
After this  transformation, the  spatial components $A_{1,2}$ remain unchanged,
whereas the time component $A_{0} = \alpha \left(1 - \rho^{2\left\vert n\right
\vert}/\left(\rho^{2\left\vert n\right\vert}+\lambda^{2\left\vert n\right\vert}
\right)\right)$ tends to zero as $\rho \rightarrow \infty$.
Hereafter, we shall use this gauge.

It is  easily  shown  that  the  Hamiltonian  (\ref{III:5})  commutes  with the
angular momentum operator
\begin{equation}
J_{3} = -i\partial_{\theta} + \sigma_{3}/2,                       \label{III:8}
\end{equation}
and that the common eigenfunctions of the operators  $H$  and  $J_{3}$ have the
form
\begin{equation}
\psi_{\varepsilon m}=\left(
\begin{array}{c}
e^{i\left(m - 1/2\right) \theta }f\left( \rho \right)  \\
e^{i\left(m + 1/2\right) \theta }g\left( \rho \right)
\end{array}%
\right) e^{-i \varepsilon t},                                     \label{III:9}
\end{equation}
where  $\varepsilon$   and   $m$  are  the  eigenvalues  of  $H$  and  $J_{3}$,
respectively.
Substitution of Eq.~(\ref{III:9}) into Eq.~(\ref{III:4}) leads to the following
system of differential equations for the radial wave functions:
\begin{eqnarray}
f^{\prime }\left( \rho \right)  &=&\rho ^{-1}\left( A_{mn}\left( \rho
\right) -1/2\right) f\left( \rho \right)                            \nonumber
  \\
&&+\left( M+\varepsilon -\alpha B_{\left\vert n\right\vert }\left( \rho
\right) \right) g\left( \rho \right),                           \label{III:10a}
  \\
g^{\prime }\left( \rho \right)  &=&\rho ^{-1}\left( -A_{mn}\left( \rho
\right) -1/2\right) g\left( \rho \right)                            \nonumber
  \\
&&+\left( M-\varepsilon +\alpha B_{\left\vert n\right\vert }\left( \rho
\right) \right) f\left( \rho \right),                           \label{III:10b}
\end{eqnarray}
where the coefficient functions are
\begin{eqnarray}
A_{mn}\left( \rho \right)  &=&m-n+\frac{n\lambda ^{2\left\vert n\right\vert }
}{\lambda ^{2\left\vert n\right\vert }+\rho ^{2\left\vert n\right\vert }}
                                                                \label{III:11a}
  \\
\text{and} \phantom{oooo} & &                                       \nonumber
  \\
B_{\left\vert n\right\vert }\left( \rho \right)  &=&\frac{\lambda
^{2\left\vert n\right\vert }}{\lambda ^{2\left\vert n\right\vert }+\rho
^{2\left\vert n\right\vert }}.                                  \label{III:11b}
\end{eqnarray}
It follows from  Eqs.~(\ref{III:10a}) -- (\ref{III:11b}) that  the  radial wave
functions $f(\rho)$ and $g(\rho)$ are real modulo  a common phase factor, since
the coefficient functions $A_{mn}(\rho)$ and $B_{\left\vert n\right\vert}(\rho)$
are real.

It is  easy  to   see  that   the   wave   function   $\psi_{\varepsilon m}$ is
an eigenfunction of the operators $P$ and $\Pi_{2} T$:
\begin{eqnarray}
\left[ \psi_{\varepsilon m n \alpha}\left(t,\mathbf{x} \right)\right]^{P}
& = & \left(-1\right)^{m-1/2}
\psi_{\varepsilon m n \alpha}\left(t,\mathbf{x}\right),         \label{III:12a}
  \\
\left[\psi_{\varepsilon m n \alpha}\left(t,\mathbf{x}\right)\right]^{\Pi_{2}T}
& = &
\psi_{\varepsilon m n \alpha}\left(t,\mathbf{x}\right),         \label{III:12b}
\end{eqnarray}
where the  parameters  $n$  and  $\alpha$  of  the soliton background field are
indicated, and the radial wave  functions  $f(\rho)$  and $g(\rho)$ are assumed
to be real.
At the same time,  the  charge   conjugation $C$  transforms  the wave function
$\psi_{\varepsilon m n \alpha}$ as follows:
\begin{equation}
\left[\psi_{\varepsilon m n \alpha}\left(t,\mathbf{x} \right) \right]^{C} =
\psi_{-\varepsilon\,-m\,-n\,-\alpha}\left(t,\mathbf{x}\right),  \label{III:14a}
\end{equation}
where in the $C$-transformed wave function
\begin{equation}
\psi_{-\varepsilon\,-m\,-n\,-\alpha}\left(t, \mathbf{x} \right) =
\left(
\begin{array}{c}
e^{i\left( -m-1/2\right) \theta }g\left( \rho \right)  \\
e^{i\left( -m+1/2\right) \theta }f\left( \rho \right)
\end{array}%
\right) e^{i\varepsilon t},                                     \label{III:14b}
\end{equation}
the real radial wave functions $f(\rho)$ and $g(\rho)$ are permuted compared to
Eq.~(\ref{III:9}).

Eq.~(\ref{III:14a}) tells us that the  charge  conjugation  transforms the wave
function of a  positive  energy state $\left\vert \varepsilon ,m \right\rangle$
into the wave function of a negative  energy  state $\left\vert -\varepsilon,-m
\right\rangle$.
In addition, the background soliton field with parameters $n$  and  $\alpha$ is
transformed into one with reversed parameters $-n$ and $-\alpha$.
Note  that   the  system   of   differential   equations   (\ref{III:10a})  and
(\ref{III:10b}) is  invariant  under the replacement $\varepsilon, m, n, \alpha
\rightarrow  -\varepsilon, -m, -n, -\alpha$  together   with   the  permutation
$f \leftrightarrow g$.
It follows from Eq.~(\ref{III:14a})  that  in  the study of the fermion-soliton
system,  it  is  sufficient  to   restrict  ourselves  to  fermionic  ($\propto
e^{- i \varepsilon t}$)    solutions,     since   the   antifermionic ($\propto
e^{i \varepsilon t}$)  solutions  are  obtained  from  the  fermionic  ones via
charge conjugation.

The system of differential equations (\ref{III:10a}) and (\ref{III:10b}) can be
reduced to a second-order differential  equation  for  the radial wave function
$f(\rho)$
\begin{equation}
f^{\prime \prime }\left( \rho \right) +P\left( \rho \right) f^{\prime
}\left(\rho \right)+Q\left(\rho \right) f\left(\rho\right) = 0,  \label{III:15}
\end{equation}
where the coefficient functions
\begin{equation}
P\left( \rho \right) =\frac{1}{\rho }+\frac{\alpha B_{\left\vert
n\right\vert }^{\prime }\left( \rho \right) }{M+\varepsilon -\alpha
B_{\left\vert n\right\vert }\left( \rho \right)},                \label{III:16}
\end{equation}
\begin{eqnarray}
Q\left( \rho \right)  &=&\left( \varepsilon -\alpha B_{\left\vert
n\right\vert }\left( \rho \right) \right) ^{2}-M^{2}                \nonumber
  \\
&&-\frac{\left( 1/2-A_{mn}\left( \rho \right) \right) ^{2}}{\rho ^{2}}-\frac{
A_{mn}^{\prime}\left(\rho \right)}{\rho}                            \nonumber
  \\
&&+\alpha \frac{1/2-A_{mn}\left( \rho \right) }{\rho }\frac{B_{\left\vert
n\right\vert }^{\prime }\left( \rho \right) }{M+\varepsilon -\alpha
B_{\left\vert n\right\vert }\left( \rho \right)},                \label{III:17}
\end{eqnarray}
and we have turned  to  dimensionless  variables  according to the substitution
rule: $\rho\rightarrow \lambda\rho$, $M\rightarrow \lambda^{-1}M$, $\varepsilon
\rightarrow \lambda^{-1} \varepsilon$, $\alpha \rightarrow \lambda^{-1}\alpha$.
To obtain the second-order differential equation  for  the radial wave function
$g(\rho)$, we must perform  the substitutions $f  \rightarrow  g,\, \varepsilon
\rightarrow -\varepsilon, m \rightarrow -m, n\rightarrow -n, \alpha \rightarrow
-\alpha$ in Eqs.~(\ref{III:15}) -- (\ref{III:17}).

The analysis shows that for the minimum possible $\left\vert n \right\vert= 2$,
differential equation (\ref{III:15})  has  nine regular singular points (one of
which is $\rho = 0$), and  one  irregular  singular  point at spatial infinity.
For $\left\vert n\right\vert >2$, the number of regular singular points is even
greater.
This means that the  solution   to  differential equation (\ref{III:15}) cannot
be expressed in terms of known special functions \cite{Slavyanov}.
Note, however, that in some  field  models, the  background field approximation
allows fermionic wave functions to be found in an analytical form.
In  particular,  it   was  found  in  Ref.~\cite{loginov_prd_2023}  that  for a
fermion-soliton system of the original $\mathbb{CP}^{N-1}$ model, the fermionic
wave functions  could  be  expressed  in  terms  of  the  local  confluent Heun
functions \cite{Ronveaux, DLMF}.
In addition, it  was  shown in Refs.~\cite{loginov_epjc_2022, loginov_npb_2022}
that the fermion scattering on  a  one-dimensional  kink  or  Q-ball could also
be described in terms of local Heun functions \cite{Ronveaux, DLMF}.

Next, we shall consider the minimal possible case $\left\vert n \right\vert=2$.
Using standard methods of  analysis,  we find that for small $\rho$, the radial
wave functions have the forms
\begin{eqnarray}
f\left( \rho \right)  &=&N\rho ^{\mu }\left( 1+\frac{M^{2}-\left(
\varepsilon -\alpha \right) ^{2}}{4(\mu +1)}\rho ^{2}+O\left[ \rho ^{4}
\right] \right),                                                \label{III:18a}
  \\
g\left( \rho \right)  &=&N\frac{\alpha -\varepsilon +M}{2\left( \mu
+1\right) }\rho ^{\mu +1}                                           \nonumber
  \\
&&\times \left( 1+\frac{M^{2}-\left( \varepsilon -\alpha \right) ^{2}}{4(\mu
+2)}\rho ^{2}+O\left[ \rho ^{4}\right] \right),                 \label{III:18b}
\end{eqnarray}
where $N$ is a normalisation  factor, $\mu = m - 1/2$, and the angular momentum
eigenvalues $m = 1/2, 3/2, 5/2, \ldots$
For  $m = -1/2, -3/2, -5/2, \ldots$, the small $\rho$ asymptotics of the radial
wave functions is
\begin{eqnarray}
f\left( \rho \right)  &=&N\frac{-\alpha +\varepsilon +M}{2\left( \mu
+1\right) }\rho ^{\mu +1}                                           \nonumber
 \\
&&\times \left( 1+\frac{M^{2}-\left( \varepsilon -\alpha \right) ^{2}}{4(\mu
+2)}\rho ^{2}+O\left[ \rho ^{4}\right] \right),                 \label{III:18c}
  \\
g\left( \rho \right)  &=&N\rho ^{\mu }\left( 1+\frac{M^{2}-\left(
\varepsilon -\alpha \right) ^{2}}{4\left( \mu +1\right) }\rho ^{2}+O\left[
\rho ^{4}\right] \right),                                       \label{III:18d}
\end{eqnarray}
where $\mu = -m - 1/2$.
We see that in both cases,  $\mu = 0, 1, 2, \ldots$,  and  therefore  plays the
role of  the  orbital  angular   momentum   and   determines  the  behaviour of
$f\left(\rho\right)$  and  $g\left(\rho\right)$ at small distances.
Furthermore, we see that in the leading order in $\rho$, the asymptotics of the
radial wave functions does not depend  on  the  sign  of  the  soliton  winding
number $n = \pm 2$.
This dependence  appears  only  in  terms   of  the  order of $\rho^{4}$ in the
parentheses in Eqs.~(\ref{III:18a}) -- (\ref{III:18d}).

It is also possible to determine the  asymptotics  of the radial wave functions
in the neighbourhood  of  the  irregular  singular  point  at spatial infinity.
Following  the  methods  described   in   Ref.~\cite{Slavyanov},  we obtain the
following asymptotics  for   the   radial   wave  functions  of  the continuous
spectrum:
\begin{eqnarray}
f_{\pm }\left( \rho \right)  &\sim &N\frac{e^{\pm ik\rho }}{\sqrt{k\rho }}
\left( 1\pm \frac{i c_{-1} }{2k\rho }+O\left[ \left( k\rho \right) ^{-2}
\right] \right),                                                \label{III:19a}
  \\
g_{\pm }\left( \rho \right)  &\sim &\pm N\frac{ik}{\varepsilon +M}\frac{
e^{\pm ik\rho }}{\sqrt{k\rho }}                                     \nonumber
  \\
&&\times \left( 1\pm \frac{i d_{-1} }{2k\rho }+O\left[ \left( k\rho \right)
^{-2}\right] \right),                                           \label{III:19b}
\end{eqnarray}
where $k^{2} = \varepsilon^{2}-M^{2}$, and the coefficients
\begin{eqnarray}
c_{-1}  &=&(m-n)(m-n-1),                                        \label{III:19c}
 \\
d_{-1}  &=&(m-n)(m-n+1).                                        \label{III:19d}
\end{eqnarray}
Similar expressions can also be obtained for the  radial  wave functions of the
discrete spectrum.

It should be noted that the long-range  gauge  field  of  the soliton decreases
sufficiently fast that there is no need to modify the pre-exponential factor in
Eqs.~(\ref{III:19a}) and (\ref{III:19b}).
It remains equal to $\left( k \rho \right)^{-1/2}$,  which is standard  for the
two-dimensional case  and allows us to correctly  determine the phase shifts of
fermionic scattering.
Furthermore, in Eqs.~(\ref{III:19a}) and (\ref{III:19b}), the  leading terms of
the asymptotic expansions  do  not  depend  on  the  parameter  $\alpha$, which
determines the  time  dependence  of the soliton solution in Eq.~(\ref{II:11}).
It can be shown that  this  dependence  appears  only  in  terms  of  the order
$(k \rho)^{-3}$.

\subsection{\label{subsec:IIIB} Fermionic bound states}

In this subsection,  we  discuss  some  issues  concerning  the fermionic bound
states.
It is convenient to perform the substitution
\begin{eqnarray}
f\left( \rho \right)  &=&\sqrt{\frac{\varepsilon +M-\alpha +\left(
\varepsilon +M\right) \rho ^{4}}{\rho \left( 1+\rho ^{4}\right) }}u\left(
\rho \right),                                                   \label{III:19e}
 \\
g\left( \rho \right)  &=&\sqrt{\frac{\varepsilon -M-\alpha +\left(
\varepsilon -M\right) \rho ^{4}}{\rho \left( 1+\rho ^{4}\right) }}v\left(
\rho \right)                                                    \label{III:19f}
\end{eqnarray}
and to obtain differential equations for  the new radial functions $u\left(\rho
\right)$ and $v\left(\rho \right)$ in the normal form
\begin{eqnarray}
u^{\prime \prime }\left( \rho \right) -\left[ \varkappa ^{2} +
U\left(\rho \right)\right] u\left(\rho\right)&=&0,              \label{III:19g}
 \\
v^{\prime \prime }\left( \rho \right) -\left[ \varkappa ^{2} +
V\left(\rho\right)\right]v\left(\rho\right)  &=&0,              \label{III:19h}
\end{eqnarray}
where the potentials
\begin{eqnarray}
U\left( \rho \right)  &=&\left( m-1\right) m\rho ^{-2}
+W\left( \rho \right),                                         \label{III:19h1}
\\
V\left( \rho \right)  &=&\left( m+1\right) m\rho ^{-2}
+\tilde{W}\left(\rho \right),                                  \label{III:19h2}
\end{eqnarray}
and the parameter $\varkappa^{2} = M^{2}-\varepsilon^{2}$.
Both $U\left(\rho\right)$ and $V\left(\rho\right)$ are  sums of the centrifugal
and interaction potentials.
The interaction  potentials  $W\left( \rho \right)$  and  $\tilde{W}\left( \rho
\right)$  are  rational  functions   of   the  radial  variable  $\rho$ and the
parameters $m$, $n$, $\alpha$, $\varepsilon$, and $M$.
An explicit form  of  the  potential $W\left(\rho, m, n, \alpha, \varepsilon, M
\right)$ is given in Appendix A.
The  potential  $\tilde{W}\left(\rho, m, n, \alpha, \varepsilon, M \right)$  is
determined from $W\left(\rho, m, n,\alpha,\varepsilon,M\right)$ by the symmetry
relation
\begin{equation}
\tilde{W}\left(\rho,m,n,\alpha,\varepsilon ,M\right) =
W\left(\rho,\!-m,\!-n,\!-\alpha,\!-\varepsilon,M\right).       \label{III:19h3}
\end{equation}

We now consider the case $\alpha = 0$ when  the time component $A_{0}$ of gauge
field (\ref{II:12}) vanishes.
The potentials $W\left( \rho \right)$  and  $\tilde{W}\left( \rho \right)$ then
cease to  depend  on the parameters $\varepsilon$ and $M$, which simplifies the
analysis.
In particular, in  the  limiting  case  $\varkappa^{2} = 0$ ($\varepsilon = \pm
M$), the solutions to Eqs.~(\ref{III:19g}) and (\ref{III:19h}) can be expressed
in terms of elementary functions.
These solutions, however, are not normalised except for the two cases $m = 1/2,
n = 2, \varepsilon = M$ and $m = -1/2, n = -2, \varepsilon = -M$.
In the first case, the normalised solution is  $u_{M\,1/2\,2}=2\pi^{-1/2}\rho^{
1/2}\left(1+\rho ^{4}\right)^{-1/2}$,  and  in  the second case, it is $v_{-M\,
-1/2\,-2} = 2\pi ^{-1/2}\rho ^{1/2}$ $\times\left( 1+\rho ^{4}\right) ^{-1/2}$.
Returning to the initial notation,  we  obtain  two  normalised  fermionic wave
functions
\begin{equation}
\psi _{M\frac{1}{2}2}=\frac{\sqrt{2}}{\pi}
\begin{pmatrix}
\left( 1+\rho ^{4}\right)^{-1/2} \\
0
\end{pmatrix}
e^{-iMt}                                                        \label{III:19i}
\end{equation}
and
\begin{equation}
\psi _{-M-\frac{1}{2}-2}=\frac{\sqrt{2}}{\pi}
\begin{pmatrix}
0 \\
\left( 1+\rho ^{4}\right)^{-1/2}
\end{pmatrix}
e^{iMt},                                                        \label{III:19j}
\end{equation}
which correspond to the so-called half-bound fermionic states.

In Eqs.~(\ref{III:19g}) and (\ref{III:19h}), the  potentials $U$ and  $V$  have
second-order poles at $\rho= 0$, and tend to zero as $\rho \rightarrow \infty$.
From the general properties  of the Schr\"{o}dinger equation \cite{Landau III},
it follows that for bound states  ($\varkappa^{2} > 0$) to exist,  both $U$ and
$V$ must take negative values.
An analysis shows that for $\alpha  =  0$,  both  $U$  and  $V$ have domains of
negative values only when $ m = 3/2,\, n = 2$ or $m = -3/2,\, n = -2$.
In all other cases, at least one of the potentials $U$ and $V$  turns out to be
positive for all  $\rho \in (0, \infty)$,  which  makes  the existence of bound
fermionic states impossible.

Consider one of the possible cases, say $m = 3/2,\, n = 2$.
It can be shown that in the  limiting  case  $\varkappa^{2} = 0$,  the solution
$u_{M\,3/2\,2}  \propto \rho^{3/2}\left(1 + \rho ^{4}\right)^{-1/2} \sim \rho^{
-1/2}$, and hence it is not normalised.
We note that Eq.~(\ref{III:19g})  admits  a mechanical analogy, as it describes
the one-dimensional motion of a unit mass particle along the $u$-axis over time
$\rho$.
The particle starts from the origin with  zero initial velocity at time $\rho =
0$, and moves along the $u$-axis under the action  of the time-dependent linear
force $F(\rho)=\left(\varkappa^{2}+U\left(\rho, 3/2, 2\right) \right) u(\rho)$.
For the trajectory of the particle  to  correspond  to  a  normalised solution,
the particle  must  tend  to  the  origin  fast  enough  ($\propto \rho^{-1/2 +
\epsilon}$) as $\rho \rightarrow \infty$.

We have shown above, however, that  for  $\varkappa^{2} = 0$, the trajectory of
the particle does not correspond to a normalised solution.
For bound states, the  parameter $\varkappa^{2} =  M^{2} - \varepsilon^{2} $ is
positive,   which   corresponds    to    an    elastic  {\it{repulsive}}  force
$\varkappa^{2} u$.
However, if  the  particle  does  not  approach  the  origin  fast  enough when
$\varkappa^{2} = 0$, it is slowed  further  or  even  change  the  direction of
movement when the additional repulsive force appears.
Hence, for $\alpha = 0$, there are no  bound fermionic  states corresponding to
$m = 3/2$, $n = 2$.
The case $m = -3/2$, $n = -2$  can be  treated similarly, with the same result.
Thus, we conclude that there are  no  bound  fermionic  states if the parameter
$\alpha = 0$.

We now consider the case when the parameter $\alpha \ne 0$.
We note that the Hamiltonian (\ref{III:5}) describes the minimal interaction of
a massive Dirac fermion with gauge field (\ref{II:12}).
This gauge field, however, is  not  dynamic,  since there is no gauge kinematic
term in the Lagrangian (\ref{II:1}).
If the parameter $\alpha = 0$, the fermion interacts with a purely ``magnetic''
field with strength $B^{\pm 2} = \pm 8\rho^{2}\left(1 + \rho^{4} \right)^{-2}$,
where the superscript indicates the winding number of the soliton.
As shown  above,  the   bound   fermionic   states   are   absent in this case.
If the parameter $\alpha  \ne  0$, then  besides  the  ``magnetic''  field, the
fermion also interacts with  an  ``electric''  field   with     radial strength
$E_{\rho}^{\pm 2} = 4 \alpha \rho^{3}\left(1 + \rho ^{4}\right)^{-2}$.
When the parameter  $\alpha > 0$,  the  ``electric''  field  is  repulsive, and
hence there are no bound fermionic states.
However, the   ``electric''   field    is   attractive   when   $\alpha  <  0$;
in this case, fermionic bound states  are  possible, and their existence can be
established by numerical methods.
The situation is reversed  for  antifermionic  bound  states:  these states may
exist if $\alpha > 0$, and  do not exist if $\alpha < 0$.

\subsection{\label{subsec:IIIC} General formulae for fermion-soliton scattering}

In this subsection, we present general  formulae  describing fermion scattering
for the two-dimensional case.
According to general principles of  the  theory of scattering \cite{Landau III,
Taylor}, for  a  fermion  with  initial  momentum  $\mathbf{k}  =  (k, 0)$, the
asymptotic form  of  the  wave  function  of  a  fermionic  scattering state is
\begin{equation}
\Psi \sim \psi _{\varepsilon ,\mathbf{k}\,}+\frac{1}{\sqrt{2\varepsilon }}
u_{\varepsilon ,\mathbf{k}^{\prime }}f\left( k,\theta \right) \frac{
e^{ik\rho }}{\sqrt{-i\rho }},                                    \label{III:20}
\end{equation}
where
\begin{equation}
\psi _{\varepsilon ,\mathbf{k}}=\frac{1}{\sqrt{2\varepsilon}}
\begin{pmatrix}
\sqrt{\varepsilon + M} \\
i\sqrt{\varepsilon - M}
\end{pmatrix}
e^{-i k x}                                                       \label{III:21}
\end{equation}
is the wave function of the incoming fermion with momentum $\mathbf{k}=(k, 0)$,
\begin{equation}
u_{\varepsilon ,\mathbf{k}^{\prime }}=
\begin{pmatrix}
\sqrt{\varepsilon + M} \\
i\sqrt{\varepsilon - M}e^{i\theta }                              \label{III:22}
\end{pmatrix}
\end{equation}
is the spinor amplitude of  the  wave  function  of  the  outgoing fermion with
momentum $\mathbf{k}^{\prime} = (k\cos(\theta), k\sin(\theta))$, and $f\left(k,
\theta \right)$ is the scattering amplitude.
Due to  the  conserved  angular  momentum  (\ref{III:8}),  we   can  expand the
scattering  amplitude  $f\left(k,  \theta  \right)$   in  terms  of the partial
scattering amplitudes $f_{m}\left(k\right)$ as
\begin{equation}
f\left( k,\theta \right) =\sum\limits_{m}f_{m}\left( k\right) e^{i\left(
m-1/2\right)\theta},                                             \label{III:23}
\end{equation}
where the summation is taken over the  half-integer  eigenvalues of the angular
momentum.

Similarly to Eq.~(\ref{III:23}),  wave  function (\ref{III:20}) for a fermionic
scattering  state  can also  be  expanded  into  partial  waves as $\Psi = \sum
\nolimits_{m}\psi_{m}$.
The large  $\rho$  asymptotics  of  the  partial  waves  $\psi_{m}$ is
\begin{equation}
\psi_{m} \sim \frac{\left(-1\right)^{1/4}}{\sqrt{2\pi k\rho }}
\begin{pmatrix}
-i\sqrt{\frac{\varepsilon + M}{2\varepsilon }}F_{m}\left( k,\rho \right)
e^{i\left( m-1/2\right) \theta } \\
\sqrt{\frac{\varepsilon -M}{2\varepsilon }}G_{m}\left( k,\rho \right)
e^{i\left( m+1/2\right) \theta }
\end{pmatrix},                                                   \label{III:24}
\end{equation}
where the radial functions
\begin{eqnarray}
F_{m}\left( k,\rho \right)  &=&i\left(-1\right)^{m-1/2}e^{-i k \rho
}+S_{m}\left(k\right)e^{ik\rho },                               \label{III:25a}
 \\
\text{and} \phantom{oooooo} & &                                     \nonumber
 \\
G_{m}\left( k,\rho \right)  &=&i\left( -1\right) ^{m+1/2}e^{-i k \rho
}+S_{m}\left(k\right)e^{ik\rho}                                 \label{III:25b}
\end{eqnarray}
are expressed in terms of the partial elements of the $S$-matrix
\begin{equation}
S_{m}\left(k\right) = 1 + i\sqrt{2 \pi k}f_{m}\left(k\right).    \label{III:26}
\end{equation}

The differential cross-section  of the elastic fermion scattering are expressed
in terms of the scattering amplitude $f\left(k, \theta \right)$ as
\begin{equation}
d\sigma/d\theta =\left\vert f\left(k, \theta \right) \right\vert^{2}.
                                                                 \label{III:27}
\end{equation}
Similarly, the   partial  cross-sections  of the elastic fermion scattering are
expressed in terms of the partial scattering amplitudes as
\begin{equation}
\sigma _{m}=2\pi \left\vert f_{m}\left( k\right) \right\vert
^{2}=k^{-1}\left\vert S_{m}\left( k\right) -1\right\vert^{2}.    \label{III:28}
\end{equation}
Both $d\sigma/d\theta$ and  $\sigma_{m}$  have  the  dimension of length, as it
should be in the two-dimensional case \cite{Landau III}.
The partial elements of the $S$-matrix  satisfy  the unitarity condition $\left
\vert S_{m}\left(k \right) \right\vert = 1$, which makes it possible to express
them in terms of the partial phase shifts $\delta_{m}$ as
\begin{equation}
S_{m}\left( k\right) = e^{2 i \delta _{m}\left( k\right)}.       \label{III:29}
\end{equation}

General formulae  for  antifermion   scattering   can  be  obtained  from those
presented in this subsection via charge conjugation (\ref{III:7c}).
In particular,  the  partial  phase   shifts   of   antifermion  scattering are
expressed in terms of those of fermion scattering as
\begin{equation}
\bar{\delta}_{m n \alpha}\left(k\right) =
\delta_{-m \,-n \,-\alpha}\left(k\right),                        \label{III:30}
\end{equation}
where the dependence of the phase shifts  on the parameters $n$ and $\alpha$ is
indicated.

We now consider the  question  of  the  symmetry  of differential cross-section
(\ref{III:27}) under the reflection $\theta \rightarrow -\theta$.
It is obvious that the  symmetry  of  the  cross-section  is  determined by the
symmetry of the scattering amplitude $f\left(k, \theta \right)$.
An analysis   of   Eqs.~(\ref{III:23})   and   (\ref{III:27})  shows   that the
differential  cross-section  is  symmetric  (even) under the reflection $\theta
\rightarrow -\theta$ if  the partial scattering amplitudes satisfy the relation
$f_{-mn\alpha}\left(k\right) = f_{m n \alpha}\left( k \right)$, while the other
possibility $f_{-m n \alpha}\left(k\right)=-f_{m n \alpha}\left( k \right)$  is
incompatible with  Eq.~(\ref{III:26})  and  unitarity condition (\ref{III:29}).

In our case, the relation $f_{-mn\alpha}\left(k\right) = f_{m n \alpha}\left( k
\right)$   is   not  satisfied,   and   therefore   differential  cross-section
(\ref{III:27}) is asymmetric under the reflection $\theta \rightarrow -\theta$.
This asymmetry is due  to  the  fact  that  the  reflection $\theta \rightarrow
-\theta$ changes  the  sign  of  the  winding  number  $n$ of $\mathbb{CP}^{1}$
soliton  (\ref{II:11}),  which,  in  turn,  changes  the  sign of the component
$A_{\theta}$ of  gauge field (\ref{II:12}).
At the  same  time, we  can  expect   that  in  the  ultrarelativistic limit $k
\rightarrow  \infty$,  the  leading  term  of  the  expansion  of  differential
cross-section (\ref{III:27}) in  $k$  is symmetric under the reflection $\theta
\rightarrow -\theta$.

\section{Fermion scattering in the Born approxima- tion}         \label{sec:IV}

In Sec.~(\ref{sec:III}),  we  established  that   the  wave  functions  of  the
fermionic scattering states  (and hence the differential cross-sections) cannot
be found in analytical form.
In view of this, it is important  to  study  the fermion scattering in the Born
approximation, which gives us  a  chance  to  obtain  an approximate analytical
description for fermion-soliton scattering.

It follows from  Eqs.~(\ref{II:1})  and  (\ref{II:3})  that  the interaction of
the fermionic isodoublet with the soliton  gauge field is described by the term
\begin{equation}
V_{\text{int}}=\bar{\psi}_{a}\gamma^{\mu}A_{\mu}\psi_{a}.          \label{IV:I}
\end{equation}
In the  background  field  approximation,  the gauge field $A_{\mu}$ is defined
by Eq.~(\ref{II:12}) and does not depend on the fermion fields $\psi_{a}$.
It follows from this and Eq.~(\ref{IV:I})  that the components of the fermionic
isodoublet $\left(\psi_{1}, \psi_{2}\right)$  interact  with the gauge field of
the $\mathbb{CP}^{1}$ soliton in the same way, and independently of each other.

Using Eq.~(\ref{IV:I}) and free-fermion  wave  functions in Eqs.~(\ref{III:21})
and (\ref{III:22}),  we  can   write   the   first-order   Born   amplitude for
fermion-soliton scattering as
\begin{equation}
f\left(\mathbf{k}^{\prime},\mathbf{k}\right) =
-\left(8\pi k\right)^{-1/2}\bar{u}
_{\varepsilon,\mathbf{k}^{\prime}}\mathbb{\gamma}^{\mu}A_{\mu}\left(
\mathbf{q}\right) u_{\varepsilon ,\mathbf{k}},                     \label{IV:2}
\end{equation}
where
\begin{equation}
A_{\mu }\left( \mathbf{q}\right) = \int A_{\mu}\left(\mathbf{x}\right)
e^{-i\mathbf{q \cdot x}}d^{2}x                                     \label{IV:3}
\end{equation}
and $\mathbf{q} = \mathbf{k}^{\prime} - \mathbf{k}$  is  the momentum transfer.
The Born amplitude in Eq.~(\ref{IV:2}) can be expressed in  terms of the Meijer
$G$-functions \cite{Prudnikov III, Mathematica}.
In particular, for the minimum possible magnitude of the soliton winding number
$\left\vert n \right\vert = 2$, the Born amplitude is
\begin{eqnarray}
f\left( \mathbf{k}^{\prime },\mathbf{k}\right)&=&- \sqrt{\pi} 2^{-5/2}
\alpha \lambda ^{2}\sqrt{k}\left[ \sqrt{\frac{\varepsilon + M}{
\varepsilon -M}}\right.                                             \nonumber
  \\
&&\left. +\sqrt{\frac{\varepsilon -M}{\varepsilon +M}}e^{-i\Delta \vartheta}
\right]\! F\left( \lambda q\right)                                  \nonumber
  \\
&&+ i \sqrt{\pi} 2^{-3/2} \tau n \lambda \sqrt{k}e^{-i\Delta \vartheta
/2}G\left( \lambda q\right),                                       \label{IV:4}
\end{eqnarray}
where the form-factor functions
\begin{eqnarray}
F\left( \lambda q\right) &=&G_{0,4}^{3,0}\left( \left( \lambda q/4\right)
^{4}\left\vert 0,\frac{1}{2},\frac{1}{2},0\right. \right),        \label{IV:5a}
 \\
G\left( \lambda q\right) &=&G_{0,4}^{3,0}\left( \left( \lambda q/4\right)
^{4}\left\vert -\frac{1}{4},\frac{1}{4},\frac{3}{4},\frac{1}{4}\right.
\right),                                                          \label{IV:5b}
\end{eqnarray}
and we use the shorthand notations:
\begin{subequations}                                               \label{IV:6}
\begin{align}
\Delta \vartheta & =\vartheta'-\vartheta, \\
\tau & =\text{sign}\left( \Delta \vartheta \right),  \\
q& =2k\sin \left( \left\vert \Delta \vartheta \right\vert /2\right).
\end{align}
\end{subequations}
The condition of applicability  of  the Born approximation can be formulated as
\begin{equation}
k \lambda \gg 1 \quad \text{and} \quad  \alpha \lambda \ll 1.      \label{IV:7}
\end{equation}
Eq.~(\ref{IV:7}) tells  us  that  the  Born  approximation  is suitable for the
scattering of high-ehergy fermions in the background field of a slowly rotating
$\mathbb{CP}^{1}$ soliton.

From Eq.~(\ref{IV:4}), it follows that
\begin{equation}
f\left( \mathbf{k}^{\prime },\mathbf{k}\right) =
f^{\ast }\left( \mathbf{k},\mathbf{k}^{\prime }\right).            \label{IV:8}
\end{equation}
We see  that  the  amplitude $f\left(\mathbf{k}^{\prime}, \mathbf{k}\right)$ is
Hermitian, as it should be  in  the  first Born approximation \cite{Landau III,
 Taylor}.
Furthermore, it can be  shown  that  the  amplitude  of  antifermion scattering
differs only   in   sign   from   the   amplitude   of   fermion  scattering in
Eq.~(\ref{IV:4}).
It follows that  in   the   first   Born   approximation,   antifermion-soliton
scattering  is  essentially  no   different   from  fermion-soliton scattering.
This also implies  that  the  scattering  of  fermions  on  a  soliton with the
parameters $(n, \alpha)$ is equivalent  to the  scattering of antifermions on a
soliton with the opposite parameters $(-n,-\alpha)$, which is in agreement with
Eq.~(\ref{III:14a}).

We now study the behaviour of  the Born amplitude for large and small values of
the momentum transfer $q$.
To do this, we use the  known  asymptotic  forms  of  the  Meijer $G$-functions
\cite{Prudnikov III, Mathematica}.
For $\lambda q \gg 1 $,  and hence $\left\vert \Delta \vartheta \right\vert \gg
\left(\lambda k \right)^{-1}$, we find that the Born amplitude
\begin{eqnarray}
f &\sim &- \pi 2^{-3/2} \left( \alpha \lambda \right) \lambda
^{1/2}e^{-\lambda q/\sqrt{2}}\left( 1+e^{-i\Delta\vartheta}\right)  \nonumber
 \\
&&\times \sin \left( \pi /8+\lambda q/\sqrt{2}\right) \sin \left( \left\vert
\Delta \vartheta \right\vert /2\right)^{-1/2}                       \nonumber
 \\
&&- i \pi 2^{-1/2} n \tau \lambda ^{1/2}e^{-\lambda q/\sqrt{2}}e^{-i\Delta
\vartheta /2}                                                       \nonumber
 \\
&&\times \cos \left( \pi /8-\lambda q/\sqrt{2}\right) \sin \left( \left\vert
\Delta \vartheta \right\vert /2\right) ^{-1/2}.                    \label{IV:9}
\end{eqnarray}
It follows from  Eq.~(\ref{IV:9}) that both the ``electric'' ($\propto \alpha$)
and ``magnetic''  ($\propto  n$)   parts   of   the   Born  amplitude  decrease
exponentially with an increase in the momentum transfer $q$.
In addition, both of these parts are oscillating functions of the dimensionless
combination $\lambda q$, due to the corresponding trigonometric factors.

Next, we consider  the  case  of   low   momentum  transfer in which $\lambda q
\rightarrow 0$, $\lambda k \gg 1$, and hence $\left\vert \Delta \vartheta\right
\vert \ll \left( \lambda k\right) ^{-1} \ll 1$.
In this case, the asymptotic form of the Born amplitude is
\begin{equation}
f \sim -(\pi/2)^{3/2}\alpha \lambda^{2}k^{1/2} - i \sqrt{2\pi} n \tau
k^{1/2}q^{-1}.                                                    \label{IV:10}
\end{equation}
We see that for the low momentum transfer  $q$,  the ``electric''  part  of the
Born amplitude does not depend on $q$, whereas  the  ``magnetic'' part diverges
$\propto q^{-1}$.
Furthermore, the magnetic part  of  amplitude  (\ref{IV:10}) does not depend on
the parameters  $\alpha$  and  $\lambda$   of   the  soliton  solution,  but is
determined only by the momentum $k$ and the scattering angle $\Delta \vartheta$
of the fermion.

We now study the partial Born  amplitudes $f_{m}\left( k \right) = \left( 2 \pi
\right)^{-1}\int\nolimits_{-\pi}^{\pi} e^{-i\left(m-1/2 \right)\Delta\vartheta}
f\left( k,\Delta \vartheta \right) d\Delta \vartheta$,  where $f\left( k,\Delta
\vartheta \right)$ is the first-order Born amplitude in Eq.~(\ref{IV:4}).
It can be shown that the imaginary part  of  the  integrand diverges as $\Delta
\vartheta \rightarrow -\pi \, \text{or} \, \pi$,  and  is  an  odd  function of
$\Delta \vartheta$.
It follows that the imaginary part of the integral vanishes in the sense of the
principal value.
In contrast, the real part  of  the  integrand  is  a  finite, even function of
$\Delta \vartheta$, meaning that the real part  of  the  integral exists and is
nonzero.
Hence, the partial amplitudes  of fermion scattering are real in the first Born
approximation.
Note that  this  property  of   the   partial   Born   amplitudes  follows from
Eqs.~(\ref{III:23}) and (\ref{IV:8}). 

In general, the partial Born amplitudes cannot be obtained  in analytical form;
however, it  is  possible  to  obtain their asymptotic  forms in the parametric
domain  $1 \lesssim \left\vert m \right\vert \ll k \lambda $ as
\begin{equation}
f_{m n}\sim 2^{-1}\sqrt{\pi }k^{-1/2}\left[ -\alpha \lambda + n m \left(k
\lambda \right) ^{-1}\right].                                   \label{IV:11a}
\end{equation}
Then, using  Eqs.~(\ref{III:26})  and  (\ref{IV:11a}), we can obtain asymptotic
forms of the partial elements of the $S$-matrix as
\begin{equation}
S_{m n} \sim 1+i 2^{-1/2}\pi\left[-\alpha \lambda + m n \left(k \lambda\right)
^{-1}\right].                                                    \label{IV:11b}
\end{equation}

It follows from Eq.~(\ref{IV:11b}) that the partaial elements of the $S$-matrix
do not satisfy the unitarity condition  in  Eq.~(\ref{III:29}),  and  this is a
characteristic property of the Born approximation \cite{Landau III, Taylor}.
We also see that the imaginary  part  of  Eq.~(\ref{IV:11b})  is much less than
unity when $\left\vert m \right\vert \ll  k \lambda$ and condition (\ref{IV:7})
is satisfied.
In this case, we can rewrite Eq.~(\ref{IV:11b}) in the approximate unitary form
$S_{m n}\approx\exp(2 i \delta_{m n}^{\text{B}})$, where the Born partial phase
shifts
\begin{equation}
\delta_{m n}^{\text{B}}=2^{-3/2}\pi \left[ -\alpha \lambda
+ m n \left( k\lambda \right) ^{-1}\right].                      \label{IV:11c}
\end{equation}
It is known  that the  condition  for validity of the Born approximation is the
smallness of the partial phase shifts \cite{Landau III, Taylor}.
Eq.~(\ref{IV:11c}) tells us that that  this condition will be met provided that
$\left\vert  m  \right\vert  \ll  k \lambda$  and condition (\ref{IV:7}) holds.

The analytical form (\ref{IV:4})  of  the first-order  Born amplitude  makes it
possible to obtain  an  expression  for  the  differential cross-section of the
fermion scattering in the Born approximation as follows:
\begin{eqnarray}
\frac{d\sigma }{d \Delta \vartheta } &=&\frac{\pi}{2} k\lambda ^{2}G^{2}
- \frac{\pi }{4}\alpha n M\lambda^{3}\sin \left(\left\vert \Delta \vartheta
\right\vert /2\right) F G                                           \nonumber
 \\
&&+\frac{\pi }{8}\alpha ^{2}\lambda ^{4}k^{-1}\left( M^{2}+k^{2}\cos^{2}\left(
\Delta  \vartheta /2\right) \right) F^{2},                        \label{IV:12}
\end{eqnarray}
where the form-factor functions $F$  and  $G$ are defined by Eqs.~(\ref{IV:5a})
and (\ref{IV:5b}), respectively.
The same expression is valid for antifermion scattering.
We  see  that  differential  cross-section  (\ref{IV:12})  is  the  sum  of the
``magnetic'' ($\propto G^2$),  interference  ($\propto F G$), and  ``electric''
($\propto F^{2}$) terms.
It is symmetric under  the  reflection $\Delta  \vartheta  \rightarrow  -\Delta
\vartheta$, which is  a  consequence  of  the Hermiticity of the Born amplitude
(\ref{IV:4}).
Furthermore, differential  cross-section  (\ref{IV:12})  is invariant under the
replacement $\alpha,\,n \rightarrow -\alpha,\,-n$.
It follows  that   in   the   first   Born   approximation,    scattering of an
(anti)fermion on a soliton  with  parameters  $\alpha$ and $n = \pm 2$ does not
differ  from  scattering  on  a  soliton  with   parameters  opposite  in sign,
$-\alpha$ and $n = \mp 2$.

In the region of small scattering  angles $\Delta \vartheta \ll \left(k \lambda
\right) ^{-1}$, the asymptotic form of differential cross-section (\ref{IV:12})
is
\begin{equation}
\frac{d\sigma }{d\Delta \vartheta }\sim \frac{32\pi \lambda }{\left(
k\lambda \right) \left( \Delta \vartheta \right) ^{2}}-8\pi ^{2}\lambda
\left( k\lambda \right) +\frac{\pi ^{3}}{2}\lambda \left( \alpha \lambda
\right) ^{2}\left( k\lambda \right),                              \label{IV:14}
\end{equation}
where the large ($k \lambda$) and  small ($\Delta \vartheta$, $\alpha \lambda$)
dimensionless parameters are indicated by parentheses.
We see that the main contribution  to the differential cross-section comes from
the ``magnetic''  term in Eq.~(\ref{IV:12}), and that it diverges quadratically
as $\Delta \vartheta \rightarrow 0$.
Hence,   the   total   cross-section   $\sigma  =  \int  \nolimits_{-\pi}^{\pi}
\left(d\sigma/d\Delta\vartheta\right) d\Delta\vartheta$  of the fermion-soliton
scattering is infinite.
However, the transport cross-section $\sigma_{\text{tr}}=\int\nolimits_{-\pi}^{
\pi}\left(1 - \cos\left( \Delta \vartheta \right) \right)\left( d\sigma/d\Delta
\vartheta\right) d\Delta \vartheta$ remains finite.
Using approximate analytical methods, it can be shown that
\begin{equation}
\sigma _{\text{tr}}\approx \frac{24.44}{\left(k \lambda\right) ^{2}}\lambda
+\frac{2.23}{\left(k \lambda\right) ^{2}}(\alpha \lambda)^{2}\lambda+O\left[
\left( \lambda k\right) ^{-4}\right].                             \label{IV:15}
\end{equation}
We see that the transport cross-section tends to zero $\propto(k \lambda)^{-2}$
as the dimensionless parameter $k \lambda \rightarrow \infty$.

\section{Numerical results}                                       \label{sec:V}

In this section, we find the dependence  of  the  energy $\varepsilon_{m n}$ of
bound (anti)fermionic states on the parameter $\alpha$.
We also find the dependence of  the  partial phase shifts $\delta_{m n}$ on the
fermion momentum.
In both cases, we restrict ourselves to the soliton winding number $n = \pm 2$,
which corresponds to the minimal allowed magnitude $\left\vert n\right\vert=2$.
To solve  these  problems,  we   use   numerical   methods  implemented  in the
{\sc{Mathematica}} software package \cite{Mathematica}.
We also  use  dimensionless  variables   according   to  the  substitution rule
$\rho \rightarrow \lambda \rho, \; M \rightarrow \lambda^{-1} M, \; \varepsilon
\rightarrow \lambda^{-1} \varepsilon, \;\alpha \rightarrow \lambda^{-1}\alpha$,
where the parameter $\lambda$  determines  the  effective  size of the soliton.
In addition, the  dimensionless  fermionic  mass  $M$  is  taken to be equal to
unity.

To determine the energy levels  of  bound  (anti)fer- mionic  states  for a given
value of $\alpha$, we must find  the  values of the parameter $\varepsilon$ for
which the solutions  to   Eq.~(\ref{III:15})  satisfy   the  necessary boundary
conditions: $f(0)$ $ = 0$ and $f(\infty) = 0$.
Note that  unlike  the  usual  eigenvalue  problem,  in Eq.~(\ref{III:15}), the
differential operator depends on the parameter $\varepsilon$.
To solve this  generalised  eigenvalue  problem,  we  used the shooting method.
Some small value of $\rho$  is  chosen  as  the initial  point  of the shooting
method, since $\rho  =  0$ is a regular singular point of differential equation
(\ref{III:15}).
Depending on the sign of $m$, the  radial wave function $f(\rho)$ and its first
derivative  $f'(\rho)$  can  be    determined    at   the   initial  point from
Eq.~(\ref{III:18a}) or Eq.~(\ref{III:18c}).
Note that the second-order differential  equation  for the radial wave function
$g(\rho)$ can also be used to find the energy of  bound (anti)fermionic states.

\begin{figure}[tbp]
\includegraphics[width=0.5\textwidth]{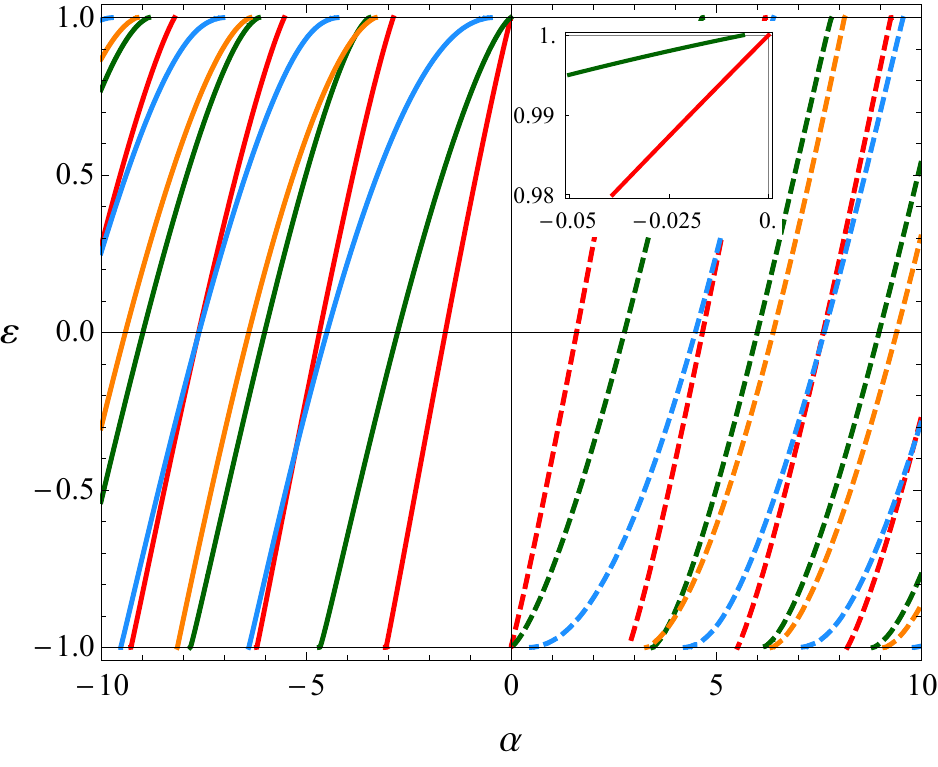}
\caption{\label{fig1}    Dependence of the energy  $\varepsilon_{m n}$ of bound
(anti)fermionic states on the  parameter $\alpha$.  The  red,  green, blue, and
orange solid curves  correspond to the fermionic levels $\varepsilon_{1/2\,2}$,
$\varepsilon_{3/2\, 2}$, $\varepsilon_{5/2\, 2}$,  and $\varepsilon_{7/2\, 2}$,
respectively.  The red, green, blue, and orange  dashed  curves  correspond  to
the antifermionic  levels  $\varepsilon_{-1/2\,-2}$, $\varepsilon_{-3/2\, -2}$,
$\varepsilon_{-5/2\, -2}$,  and $\varepsilon_{-7/2\, -2}$, respectively}
\end{figure}

\begin{figure}[tbp]
\includegraphics[width=0.5\textwidth]{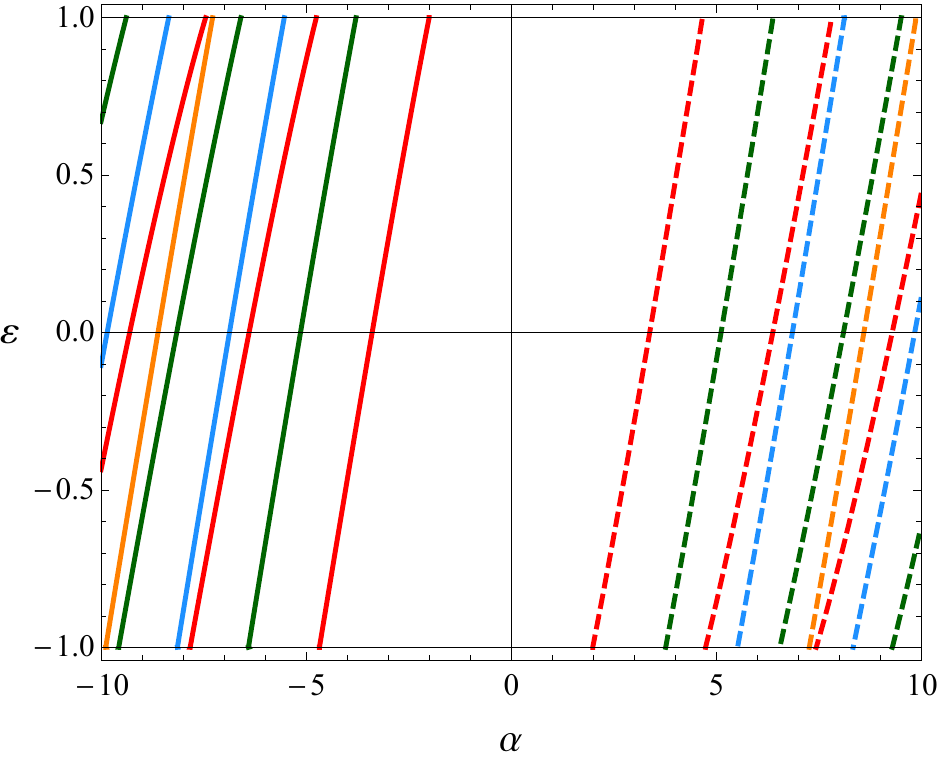}
\caption{\label{fig2}    Dependence of the energy  $\varepsilon_{m n}$ of bound
(anti)fermionic states on the  parameter $\alpha$.  The  red,  green, blue, and
orange solid curves correspond to the fermionic levels $\varepsilon_{-1/2\,2}$,
$\varepsilon_{-3/2\,2}$, $\varepsilon_{-5/2\,2}$,  and $\varepsilon_{-7/2\,2}$,
respectively.  The red, green, blue, and orange  dashed  curves  correspond  to
the  antifermionic  levels  $\varepsilon_{1/2\, -2}$, $\varepsilon_{3/2\, -2}$,
$\varepsilon_{5/2\, -2}$, and $\varepsilon_{7/2\, -2}$, respectively}
\end{figure}

\begin{figure}[tbp]
\includegraphics[width=0.5\textwidth]{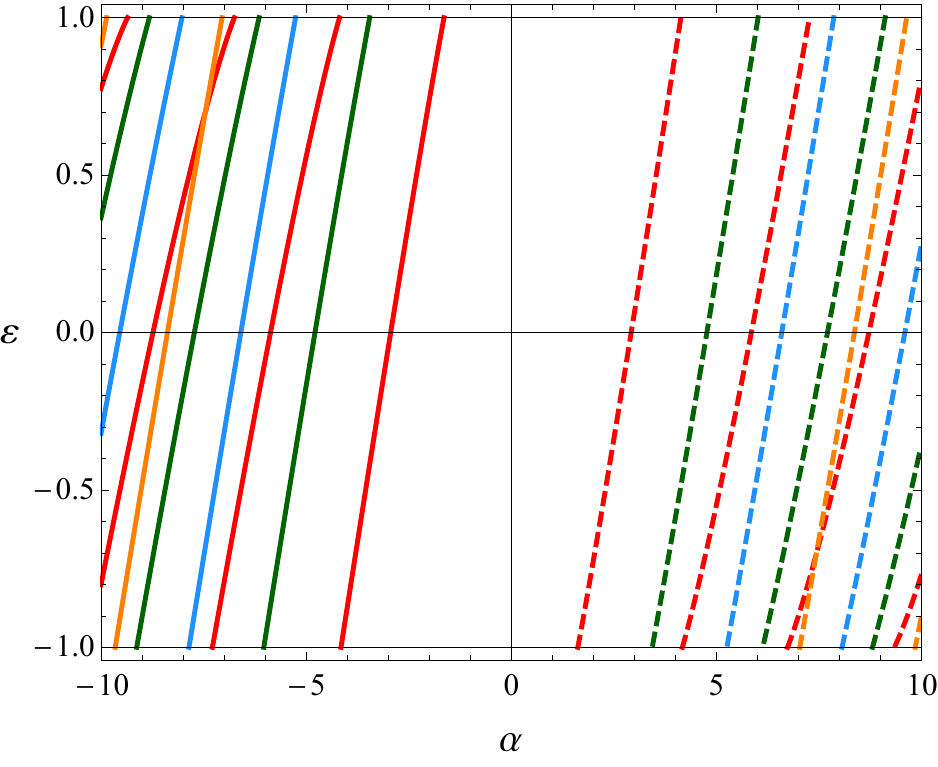}
\caption{\label{fig3}    Dependence of the energy  $\varepsilon_{m n}$ of bound
(anti)fermionic states on the  parameter $\alpha$.  The  red,  green, blue, and
orange solid curves correspond to the fermionic levels $\varepsilon_{1/2\,-2}$,
$\varepsilon_{3/2\,-2}$, $\varepsilon_{5/2\,-2}$, and $\varepsilon_{7/2\, -2}$,
respectively.  The red, green, blue, and orange  dashed  curves  correspond  to
the  antifermionic  levels  $\varepsilon_{-1/2\, 2}$, $\varepsilon_{-3/2\, 2}$,
$\varepsilon_{-5/2\, 2}$,  and $\varepsilon_{-7/2\, 2}$, respectively}
\end{figure}

\begin{figure}[tbp]
\includegraphics[width=0.5\textwidth]{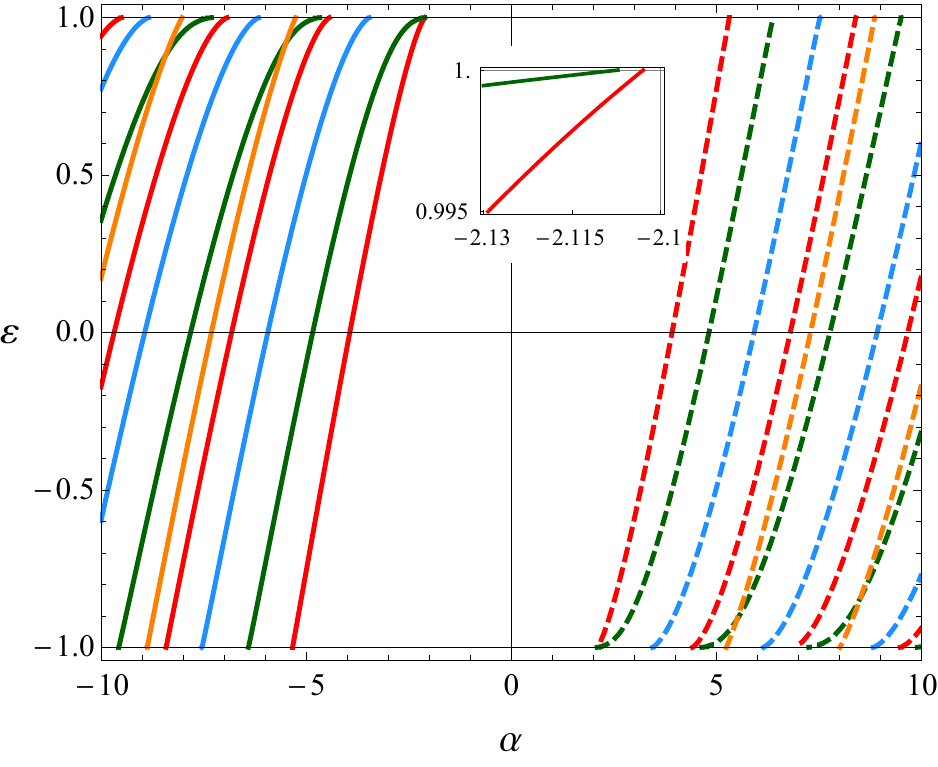}
\caption{\label{fig4}    Dependence of the energy  $\varepsilon_{m n}$ of bound
(anti)fermionic states on the  parameter $\alpha$.  The  red,  green, blue, and
orange solid curves correspond  to  the fermionic levels $\varepsilon_{-1/2\,-2
}$, $\varepsilon_{-3/2\,-2}$, $\varepsilon_{-5/2\,-2}$, and  $\varepsilon_{-7/2
\,-2}$, respectively. The red, green, blue, and orange dashed curves correspond
to the antifermionic  levels  $\varepsilon_{1/2\, 2}$, $\varepsilon_{3/2\, 2}$,
$\varepsilon_{5/2\, 2}$,  and $\varepsilon_{7/2\, 2}$, respectively}
\end{figure}

Figures~\ref{fig1} -- \ref{fig4}    show    the   dependence   of   the  energy
$\varepsilon_{m n}$  of bound (anti)fermionic states on the parameter $\alpha$.
These four figures correspond to the four possible combinations of signs of the
quantum numbers $m$ and $n$.
It follows from these figures  that  bound  fermionic  states are possible only
when the parameter $\alpha < 0$.
In contrast, bound  antifermionic  states  are possible only when the parameter
$\alpha > 0$.
Furthermore, in  Figs.~\ref{fig1} -- \ref{fig4}, all the curves $\varepsilon_{m
n}(\alpha)$ satisfy the symmetry relation
\begin{equation}
\varepsilon_{mn}\left( \alpha \right) = -\varepsilon _{-m-n}\left(-\alpha
\right),                                                            \label{V:1}
\end{equation}
which is a  consequence  of the symmetry  (\ref{III:14a}) of the Dirac equation
(\ref{III:4}) with respect to charge conjugation.
Eq.~(\ref{V:1}) tells  us  that  it  is  sufficient  to  restrict  ourselves to
negative $\alpha$  (fermionic bound  states)  to  study  the  behaviour  of the
$\varepsilon_{m n}\left( \alpha \right)$ curves.

Firstly, we note  that  in Figs.~\ref{fig1} -- \ref{fig4}, the behaviour of all
the  curves  $\varepsilon_{m n} (\alpha)$  fits a common pattern.
For each $(m,n)$, the curve $\varepsilon_{m n}(\alpha)$ consists of an infinite
sequence of branches $\varepsilon_{m n}^{(i)}(\alpha)$, $i = 1,\ldots, \infty$.
For each $(m,n)$ there exists a minimum value of $\left\vert \alpha \right\vert
=-\alpha$ above which the corresponding  bound fermionic state emerges from the
positive energy continuum for the first time.
The minimum value is positive for all $(m, n)$  except  for the fermionic state
with $(m, n) = (1/2, 2)$, for which it is equal to zero.
This is  obviously  caused  by  the  presence  of  half-bound  fermionic  state
(\ref{III:19i}) with $(m, n) = (1/2, 2)$.
It follows from the subplot in  Fig.~\ref{fig1},  that  the  first  bound state
$(3/2,2)$ emerges at a  positive  (albeit  small)  value  of  $\left\vert\alpha
\right\vert$.
It was found numerically that for  $\left\vert  m \right\vert \gtrsim 9/2$, the
first emergence of a bound fermionic  state becomes equidistant with respect to
$\alpha$.

Consider a fermionic curve $\varepsilon_{m n} (\alpha)$.
After the emergence of the  first branch $\varepsilon_{m n}^{(1)}(\alpha)$ from
the positive energy  continuum,  the  energy  $\varepsilon_{m n}$  of the bound
fermionic state decreases  monotonically with an increase in $\left\vert \alpha
\right\vert$.
With further  growth  of  $\left\vert  \alpha  \right\vert$,  the  first branch
$\varepsilon_{m n}^{(1)}(\alpha)$   reaches    the    lower   admissible  bound
$\varepsilon = -M$ and is interrupted. 
At the same time, the  second  branch $\varepsilon_{m n}^{(2)}(\alpha)$ emerges
from the positive energy continuum, and its behaviour is similar to that of the
first branch as $\left\vert\alpha\right\vert$ grows.
With  further  growth  in  $\left\vert  \alpha  \right\vert$,  this  pattern is
repeated over and over again,  resulting  in  an  infinite sequence of branches
$\varepsilon_{m n}^{(i)}(\alpha)$ for each $(m, n)$.
We find  that  for  the  $i$-th   branch $\varepsilon_{m n}^{(i)}(\alpha)$, the
radial wave function $f(\rho)$ has exactly $i - 1$ nodes.
It follows that for a given $(n, m)$,  the  branch  number  $i$  determines the
radial quantum number $n_{\rho} = i - 1$.
We also find that for a given $(m,n)$ and any fixed $\alpha$, there are at most
two  bound  (anti)fermionic states,  and their radial quantum numbers differ by
one.
The distance between neighboring branches $\varepsilon_{m n}^{(i + 1)}(\alpha)$
and $\varepsilon_{m n}^{(i)}(\alpha)$  is  approximately constant, and does not
depend on $m$, $n$, and $i$.

It follows from Figs.~\ref{fig1} -- \ref{fig4} that each branch $\varepsilon_{m
n}^{(i)}(\alpha)$ crosses the level $\varepsilon = 0$ at some nonzero $\alpha$.
This corresponds to  the  appearance  of  an (anti)fermionic  zero  mode in the
background field of the $\mathbb{CP}^{1}$ soliton.
In each of Figs.~\ref{fig1} -- \ref{fig4}, the fermionic and antifermionic zero
modes are arranged symmetrically with respect to the origin, in accordance with
Eq.~(\ref{V:1}).
Furthermore, we see that  for  a  given value of $\alpha$, there exists at most
one (anti)fermionic zero mode.

We now  proceed  to  a  study  of  the  phase  shifts  of fermionic scattering.
To find the phase shifts, we need to find a numerical solution to the system of
differential  equations  in  Eqs.~(\ref{III:10a})  and  (\ref{III:10b})  on the
interval $\left[\rho_{\min},\rho_{\max} \right]$, where $\rho_{\min} \ll 1$ and
$\rho_{\max} \gg 1$.
Since $\rho=0$ is a regular singular point of the system, we cannot set $\rho_{
\min}$ equal to zero; instead,  we  set  $\rho_{\min}$ equal to a value on  the
order  of $10^{-2}$, and  determine  the  values  of the  radial wave functions
$f(\rho)$ and        $g(\rho)$           at           $\rho_{\min}$       using
Eqs.~(\ref{III:18a}) -- (\ref{III:18d}).
We  also set  $\rho_{\max}$  equal  to  a  value  on   the  order  of $10^{2}$.
For such large  distances,  we neglect  the  terms  in Eqs.~(\ref{III:10a}) and
(\ref{III:10b}) that  decrease   faster   than   $\rho^{-1}$,   and   obtain an
approximate solution in terms of the cylindrical functions
\begin{eqnarray}
f\left( \rho \right)  &\approx &A\left( c_{1}J_{n-m+1/2}(k\rho
)+c_{2}Y_{n-m+1/2}(k\rho )\right),                                 \label{V:1a}
  \\
g\left( \rho \right)  &\approx &B\left( c_{1}J_{n-m-1/2}(k\rho
)+c_{2}Y_{n-m-1/2}(k\rho )\right),                                 \label{V:1b}
\end{eqnarray}
where the factors $A = \left( \varepsilon + M \right) ^{1/2}\left(2 \varepsilon
\right)^{-1/2}$ and $B = \left( \varepsilon -M\right) ^{1/2}\left( 2\varepsilon
\right) ^{-1/2}$.
To determine the coefficients  $c_{1}$ and $c_{2}$,  we  fit  the numerical and
approximate solutions at $\rho = \rho_{\max}$.
Using  known  asymptotic  expansions  of the cylindrical  functions and general
expressions (\ref{III:25a}) and (\ref{III:25b})  for  the radial wave functions
of a fermionic scattering state,  we  obtain  two  independent linear equations
that determine one $S$-matrix partial element $S_{m n}=\exp(2 i \delta_{m n})$.
The coincidence of the  solutions  to  these  two equations (within a numerical
error) was used  as  a  criterion  for  the  correctness  of the result.

Eq.~(\ref{A:6}) tells  us  that  as the fermion momentum $k\rightarrow \infty$,
the phase shift $\delta_{m n} \left( k \right)$  tends  to  a constant value of
$-2^{-3/2}\pi \alpha$ up  to  a  term  that  is  an  integer multiple of $\pi$.
From the  numerical  results,  we  also  find  that  as  $k \rightarrow 0$, the
fermionic phase shifts
\begin{equation}
\delta_{m n} \left( k\right) \sim a_{m n} k^{\tau},                 \label{V:2}
\end{equation}
where $a_{m n}$ is a constant, the exponent
\begin{equation}
\tau  = 2 \left\vert n - m + 1/2\right\vert,                        \label{V:3}
\end{equation}
and it is assumed that $\tau > 0$.
If $(m, n)  =  (5/2, 2)$ or $(-3/2, -2)$,  then  the  exponent $\tau$ vanishes.
We find that in these exceptional cases, the fermionic phase shifts
\begin{equation}
\delta_{m n} \left(k\right) \sim  \pi/\ln(k^{2})                    \label{V:4}
\end{equation}
as $k \rightarrow 0$.
We see that as $k \rightarrow 0$, the phase shifts tend  to zero according to a
power law when $\tau > 0$, but only logarithmically when $\tau = 0$.
This significant difference in the behaviour of  the phase shifts is due to the
fact that in  Eq.~(\ref{V:1a}), in the limit of small $k$, the Bessel  function
of  the second  kind $Y_{n - m + 1/2}(k\rho)$ diverges $\propto k^{-\tau/2}$ if
$\tau > 0$, whereas it diverges only $\propto \ln(k)$ if $\tau = 0$.
Note that the other Bessel function of the second kind $Y_{n - m - 1/2}(k\rho)$
included in Eq.~(\ref{V:1b}) does not  cause  the  logarithmic behaviour of any
fermionic phase shift as $k \rightarrow 0$, since  the  factor $B\rightarrow 0$
in this case.
Instead, the Bessel function  $Y_{n - m - 1/2}(k \rho)$  causes the logarithmic
behaviour of the antifermionic  phase shifts $\delta_{3/2\,2}(k)$ and $\delta_{
-5/2\,-2}(k)$ in the limit of small $k$.

We investigated the behaviour of  the curves $\delta_{m n}(k)$ for all possible
combinations of signs of $m$ and $n$.
In particular, we found the  limiting  values  of  the  fermionic  phase shifts
$\delta_{m n}(k)$ as $k \rightarrow \infty$ as
\begin{equation}
\delta _{mn}\left( \infty \right) = -2^{-3/2}\pi \alpha - \left( \nu
_{mn}\left( \alpha \right) + 1 \right) \pi                          \label{V:5}
\end{equation}
if $m n > 0$, and
\begin{equation}
\delta _{mn}\left( \infty \right) =-2^{-3/2}\pi \alpha -\left( \nu
_{mn}\left( \alpha \right) -1\right) \pi                            \label{V:6}
\end{equation}
if $m n < 0$.
In Eqs.~(\ref{V:5}) and (\ref{V:6}), $\nu_{m n}(\alpha)$ is the total number of
branches $\varepsilon_{m n}^{(i)}$,  where  $i = 1, \ldots, \nu_{m n}(\alpha)$,
emerging from  the  positive energy  continuum on the interval $\left[\alpha, 0
\right]$.
Eqs.~(\ref{V:5}) and (\ref{V:6}) are valid for all  $(m, n)$ except those equal
to $(-1/2,-2)$, $(1/2,2)$, and $(3/2,2)$.
For these states, the following relations hold:
\begin{eqnarray}
\delta _{-1/2\,-2}\left( \infty \right)  &=&-2^{-3/2}\pi \alpha -\nu
_{-1/2\,-2}\left( \alpha \right) \pi,                              \label{V:7a}
 \\
\delta _{1/2\,2}\left( \infty \right)  &=&-2^{-3/2}\pi \alpha -\left( \nu
_{1/2\,2}\left( \alpha \right) -1\right) \pi,                      \label{V:7b}
 \\
\delta _{3/2\,2}\left( \infty \right)  &=&-2^{-3/2}\pi \alpha -\nu
_{3/2\,2}\left( \alpha \right) \pi.                                \label{V:7c}
\end{eqnarray}
We note that the exceptional states  of  Eqs.~(\ref{V:7a}) -- (\ref{V:7c}) form
the sequences $(-3/2,-2)$, $(-1/2,-2)$ and $(1/2,2)$, $(3/2,2)$, $(5/2,2)$ with
the ``logarithmic''  states  of  Eq.~(\ref{V:4}).

It follows from  Eqs.~(\ref{V:5}) -- (\ref{V:7c}) that  the  dependence  of the
value of $\delta_{m n}(\infty)$ on  the  parameter  $\alpha$  is the sum of two
terms: the first is a regular linear function of  $\alpha$, while the second is
an irregular stepwise function of $\alpha$.
When $\alpha > 0$, the fermionic bound states  are absent, the function $\nu_{m
n}(\alpha) = 0$, and $\delta_{m n}(\infty)$ decreases linearly with an increase
in $\alpha$.
In contrast, for negative $\alpha$, the  function $\nu_{m n}(\alpha)$ increases
stepwise by $\pi$ with an increase in $\left\vert \alpha \right\vert$.
We find that the  width of the step  (the distance between two successive jumps
of $\nu_{m n}(\alpha)$) $\Delta \alpha \approx  2^{3/2} \approx 2.83$, and this
is independent on $(m, n)$.
This value of  $\Delta \alpha$  results  in the fact that for $\alpha < 0$, the
linear  term  $-2^{-3/2}  \pi  \alpha$  and  the  term $-\nu_{m n}(\alpha) \pi$
compensate each other on average, and  the dependence of $\delta_{m n}(\infty)$
on $\alpha$ therefore has a sawtooth shape.

Thus $\delta_{mn}(\infty)$   considered   as  a  function  of  $\alpha$ changes
discontinuously by $\pi$  whenever  a new bound fermionic state $(m,n)$ emerges
from the positive energy continuum.
This behaviour can be  explained  based  on  the  analytical  properties of the
$S$-matrix \cite{Taylor}.
Any partial element $S_{mn}$  of  the  $S$-matrix  can  be regarded either as a
function of the momentum $k$ or  as  a  function  of  the energy $\varepsilon =
[k^{2}+M^{2}]^{1/2}$.
In the latter case, the partial element $S_{mn}\left( \varepsilon\right)$ is an
analytical function defined on a  two-sheeted  Riemann  surface with two branch
cuts, $\left(-\infty, -M \right]$ and $\left[M, +\infty\right)$.
The physical region  of  fermionic  scattering ($k>0$) corresponds to the upper
edge of the first  (physical)  sheet  along  the  branch  cut $\left[M, +\infty
\right)$.
The  bound  fermionic  states  correspond  to  the  poles  of  $S_{n m}$, which
lie on the physical sheet in the interval $\left(-M, M\right)$.

Suppose a new bound fermionic  state $(m, n)$ emerges at $\alpha = \alpha_{0}$.
In a  small  neighbourhood  of  $\alpha_{0}$,  the  parameter  $\alpha$  can be
written as the sum of $\alpha_{0}$ and a small term $\tilde{\alpha}$: $\alpha =
\alpha_{0} + \tilde{\alpha}$.
A small positive  $\tilde{\alpha}$  corresponds  to the situation preceding the
appearance of the bound fermionic state $(m, n)$.
The theory of scattering tells us that in this case, $S_{m n}(\varepsilon)$ has
a first-order pole on the second (unphysical) sheet at the point
\begin{equation}
\varepsilon = M + \tilde{\varepsilon}_{m n} \left( \tilde{\alpha }\right) -
i\Gamma_{mn}\left(\tilde{\alpha }\right)/2,                        \label{V:7d}
\end{equation}
where   both    $\tilde{\varepsilon}_{m n}  \left(\tilde{\alpha} \right)$   and
$\Gamma_{mn}\left(\tilde{\alpha} \right)$  are  positive  and  tend  to zero as
$\tilde{\alpha} \rightarrow 0$.
In addition, $\Gamma_{mn}\left(\tilde{\alpha}\right)\ll\tilde{\varepsilon}_{mn}
\left(\tilde{\alpha}\right)$ for small positive $\tilde{\alpha}$.
Thus, for small positive $\tilde{\alpha}$, the  partial element $S_{m n}$ has a
first-order pole on the unphysical sheet under the branch cut $\left[M, +\infty
\right)$ in the small neighbourhood of the point $\varepsilon = M$.
This information  is  sufficient  to  obtain  the  Breit-Wigner formula for the
partial phase shift in the threshold region
\begin{equation}
\sin \left( \delta _{mn}
\right) =\frac{\Gamma _{mn}\left( \tilde{\alpha }\right) /2}{\bigl[
\left( \varepsilon -\varepsilon _{mn}^{r}\left( \tilde{\alpha }\right)
\right) ^{2}+\left( \Gamma _{mn}\left( \tilde{\alpha }\right) /2\right)
^{2}\bigr] ^{\frac{1}{2}}},                                        \label{V:7e}
\end{equation}
where $\varepsilon_{m n}^{r}\left(\tilde{\alpha}\right)=M + \tilde{\varepsilon}
_{m n}\left(\tilde{\alpha}\right)$.

It follows from Eq.~(\ref{V:7e})  that  if  $\alpha$  is  slightly greater than
$\alpha_{0}$, the phase shift $\delta_{m n}$  varies  from approximately $0$ to
$\pi$ in a narrow range of energies centered at the resonance point $\varepsilon
= \varepsilon_{m n}^{r}\left(\tilde{\alpha}\right)$.
However, as soon as $\alpha$  becomes  slightly less  than  $\alpha_{0}$, a new
bound fermionic state $(m,n)$ appears,  and the pole of $S_{mn}$ moves from the
old position located under the branch cut $\left[M,+\infty\right)$, through the
branch point $\varepsilon = M$, to a new position located on the physical sheet
in the interval $(-M, M)$.
The new position $\varepsilon = M - \epsilon$  of  the pole of $S_{m n}$  makes
resonant behaviour (\ref{V:7e}) impossible,  and  there  is  no  phase  jump by
$\pi$ in this case.
It follows that  $\delta_{m n}(\infty)$  decreases  discontinuously by $\pi$ as
$\alpha$ changes from  $\alpha_{0} + \epsilon$  to  $\alpha_{0} - \epsilon$, in
accordance with Eqs.~(\ref{V:5}) -- (\ref{V:7c}).

\begin{figure}[tbp]
\includegraphics[width=0.5\textwidth]{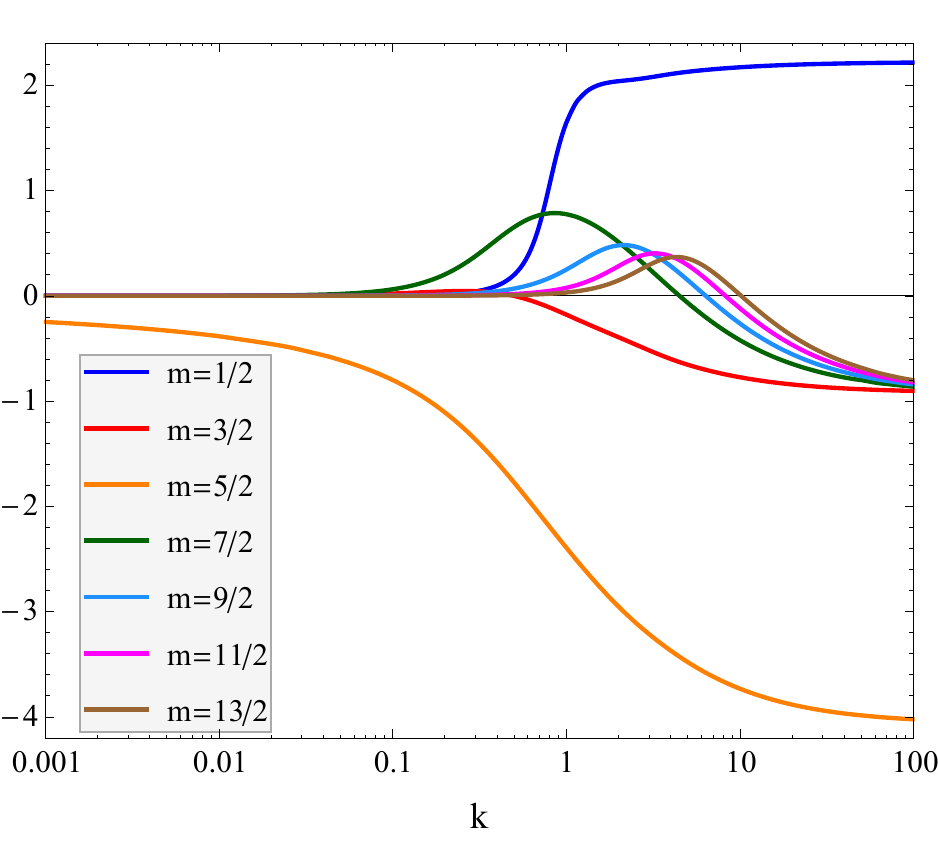}
\caption{\label{fig5}   Dependence of the fermionic phase shifts $\delta_{m 2}$
on the momentum $k$ for the  first  few  positive  values of $m$ (the parameter
$\alpha = -2$)}
\end{figure}

When presenting numerical results  for  the  phase shifts $\delta_{m n}(k)$, we
restrict ourselves to the states with $m > 0$ and $n = 2$.
For the other combinations of signs of $m$  and $n$, the behaviour of the phase
shifts $\delta_{m n}(k)$ is similar to that for  the  case considered here, and
does not provide any new information.
Figure~\ref{fig5} shows the curves $\delta_{m 2}(k)$ for the first few positive
values of $m$.
The curves correspond to the parameter $\alpha = -2$.
It follows from Fig.~\ref{fig1} that  for $\alpha = -2$, there  are three bound
fermionic states: $(1/2,2)$, $(3/2,2)$, and $(5/2,2)$.
Then, Eqs.~(\ref{V:7b}),  (\ref{V:7c}),  and  (\ref{V:5})  tell us that for the
fermionic states  $(1/2, 2)$, $(3/2, 2)$,  and  $(5/2, 2)$, the limiting values
$\delta_{m 2}(\infty)$  of  the  phase  shifts are $2^{-1/2} \pi \approx 2.22$,
$(2^{-1/2} - 1) \pi \approx -0.92$,  and  $(2^{-1/2} - 2)  \pi  \approx -4.06$,
respectively.
This is  consistent  with   the   behaviour   of   the  corresponding curves in
Fig.~\ref{fig5}.
Furthermore, it follows from Fig.~\ref{fig1}  that  for $\alpha = 2$, there are
no fermionic bound states $(m, 2)$ with $m > 5/2$.
Eq.~(\ref{V:5})  tells  us   that   for  these  states, $\delta_{m 2}(\infty) =
(2^{-1/2}-1)\pi\approx -0.92$,  which  is also consistent with Fig.~\ref{fig5}.

In Fig.~\ref{fig5}, all the curves $\delta_{m 2}(k)$ except for $\delta_{5/2\,2
}(k)$  tend  rapidly  to zero (according  to the power  law in Eq.~(\ref{V:2}))
as $k \rightarrow 0$.
In contrast, the curve $\delta_{5/2\,2}(k)$  tends  slowly to zero according to
the logarithmic law in Eq.~(\ref{V:4}).
We can rewrite  partial  cross-sections  (\ref{III:28})  in  terms of the phase
shifts as $\sigma_{mn}=4k^{-1}\sin\left(\delta_{mn}\left(k \right)\right)^{2}$.
It then follows from Eq.~(\ref{V:2}) that the partial cross-section
\begin{equation}
\sigma_{m n}(k) \propto k^{2\tau - 1} \underset{k \rightarrow 0}
{\longrightarrow} 0                                                 \label{V:8}
\end{equation}
for all $(m,n)$ except for $(5/2, 2)$ and $(-3/2,-2)$.
For these two states, the partial cross-section
\begin{equation}
\sigma_{mn}(k)\approx 2\pi^{2}k^{-1}\ln\left(k\right)^{-2},         \label{V:9}
\end{equation}
and diverges as $k \rightarrow 0$.
Hence, the main contribution  to  the  low-energy fermion scattering comes from
the ``logarithmic'' states $(5/2, 2)$ or $(-3/2,-2)$.

Eqs.~~(\ref{V:5}) -- (\ref{V:7c}) tell  us  that in the general case, the phase
shift $\delta_{m n}\left(\infty\right) \neq 0 \; \text{mod} \; \pi$.
It follows  that  the  partial  cross-sections  $\sigma_{m n}(k)$  tend to zero
$\propto k^{-1}$ as $k \rightarrow \infty$.
It also follows from  Eqs.~~(\ref{V:5}) -- (\ref{V:7c}) that for each $(m, n)$,
there exists  a  discrete  set  of $\alpha_{p} = 2^{3/2} p,\; p \in \mathbb{Z}$
for which the limiting value $\delta_{m n}\left(\infty\right) = 0 \; \text{mod}
\;\pi$.
From Eq.~(\ref{A:6}), we  see  that  if $\delta_{m n} \left(\infty\right) = 0\;
\text{mod}\;\pi$,  then  the  partial  cross-sections $\sigma_{m n}(k)$ tend to
zero $\propto k^{-3}$ as $k \rightarrow \infty$.
Note that  although  the  partial   cross-sections   tend   to  zero, the total
cross-sections diverge in  both  cases, which agrees with the results presented
in Sec.~\ref{sec:IV}.

Eqs.~(\ref{V:5}) -- (\ref{V:7c})  can  be  considered  as  a  generalisation of
Levinson's theorem \cite{levinson_49} for our relativistic case.
This theorem relates  the  difference  in  the partial phase shifts $\delta_{l}
(\infty) - \delta_{l}(0)$ to the number  of  corresponding bound states $n_{l}$
in the case of nonrelativistic potential scattering:
\begin{equation}
\delta_{l}\left(\infty\right)-\delta_{l}\left(0\right)=-\pi n_{l}. \label{V:7f}
\end{equation}
In the derivation of Eq.~(\ref{V:7f}), the nonrelativistic potential is assumed
to satisfy certain requirements \cite{Taylor}; in  particular, it must decrease
faster than  $r^{-d}$  as  $r \rightarrow \infty$,  where  $d$  is  the spatial
dimension.

The   main    difference     between     Eqs.~(\ref{V:5}) -- (\ref{V:7c})   and
Eq.~(\ref{V:7f}) is  that  in  our  case,  the  phase  shift  difference is not
equal to $0\!\! \mod \pi$ as in Eq.~(\ref{V:7f}), but is equal to $-2^{-3/2}\pi
\alpha \!\! \mod  \pi$.
It follows that the  partial elements $S_{m n}\left( k \right) = \exp \left(2 i
\delta_{m n}\left( k \right) \right)$  tend  to  the  universal  limit $\exp(-i
2^{-1/2} \pi \alpha)$  as $k \rightarrow \infty$.
This limit is  not equal to  unity,  provided that $\alpha  \ne  2^{3/2} p,\; p
\in \mathbb{Z}$,  and   we  can  therefore  say  that  in the general case, the
fermion-soliton interaction does not vanish from  the viewpoint of unitarity in
the ultrarelativistic limit $k \rightarrow \infty$.
The occurrence of  the  term $-2^{-3/2} \pi \alpha$  is due to the relativistic
character of fermionic scattering, and is explained in Appendix A.

Another difference  is  that  in  Eq.~(\ref{V:7f}), $n_{l}$  is  the  number of
really existing  bound   states   of    angular    momentum   $l$   for a given
nonrelativistic potential.
In contrast,  in   Eqs.~(\ref{V:5}) -- (\ref{V:7c}),   the   stepwise  function
$\nu_{m n}(\alpha)$ is the  total  number  of   bound fermionic $(m, n)$ states
emerging from the positive  energy  continuum  on the interval $\left[\alpha, 0
\right]$.
From Figs.~\ref{fig1} -- \ref{fig4}, we know that  for a given value of $\alpha
< 0$, some bound fermionic states can reach the lower bound $\varepsilon  = -M$
and then disappear.
Hence, for a given $\alpha$, the difference  $\delta_{m n}\left(\infty\right) -
\delta_{m n}\left( 0 \right)$ is  determined  by   both the really existing and
disappeared bound fermionic $(m, n)$ states.

\section{Conclusion}                                             \label{sec:VI}

In this paper, we have studied the  scattering  of  fermions in  the background
field of  a  topological   soliton   of   a  modified  $\mathbb{CP}^{1}$  model
\cite{leese_npb_1991}, in which a potential term was added to the Lagrangian of
the original $\mathbb{CP}^{1}$  model.
This potential  term  breaks  the  invariance  of  the  action  of the original
$\mathbb{CP}^{1}$ model under the scale transformations $\mathbf{x} \rightarrow
\lambda \mathbf{x}$.
As a result, the energy  of the soliton of the modified $\mathbb{CP}^{1}$ model
depends on its size, which is fixed by a conserved Noether charge.
For this reason, the soliton has no dilatation zero mode,  meaning that it does
not suffer from the  rolling  scale  instabilities  inherent to solitons of the
original $\mathbb{CP}^{1}$ model.

Both the original  and  modified $\mathbb{CP}^{1}$  models  are invariant under
local $U(1)$ transformations, and hence include an Abelian gauge field.
This field, however, is  not  dynamic, since there is  no corresponding kinetic
term in the Lagrangian (\ref{II:1}).
The incorporation  of fermions  into  the  $\mathbb{CP}^{1}$  model is realised
through their minimal interaction with the Abelian gauge field.
As a result, the fermion-soliton  interaction looks like an interaction between
an electrically charged  particle  and a two-dimensional  object (soliton) with
long-range electric and magnetic fields.

The presence of the  long-range  ``electric''  field  leads to the existence of
bound (anti)fermionic states.
These bound fermionic (antifermionic) states  exist only at negative (positive)
values of the parameter  $\alpha$  that  determines  the phase frequency of the
$\mathbb{CP}^{1}$ soliton.
As $\left\vert  \alpha  \right\vert$  increases,  the fermionic (antifermionic)
bound states emerge from the positive (negative) energy continuum of states and
then disappear  when  the  negative  (positive)  energy  continuum of states is
reached.
The invariance  of  the  model  under  charge  conjugation  leads  to a certain
symmetry between the fermionic and antifermionic bound states.

In addition to the bound (anti)fermionic states of the discrete spectrum, there
exist scattering (anti)fer- mionic states of the continuous spectrum.
We have  investigated  the  (anti)fermion  scattering  in  the framework of the
Born and semiclassical approximations, and also by numerical methods.
In particular, we  have  found  that  due  to  the  long-range character of the
Abelian gauge field, the total scattering  cross-section  diverges, whereas the
transport cross-section remains finite.

Fermion scattering can  be  completely  described  in  terms  of  partial phase
shifts.
In view of this, we have  investigated  the  momentum dependence of the partial
phase shifts using approximate analytical and numerical methods.
In particular, we have established the relations between  the difference in the
partial phase shifts and  the  number  of corresponding bound fermionic states.
These relations are a generalisation of Levinson's theorem for our relativistic
case.
The main difference is  that  in  our case, the difference in the partial phase
shifts is not equal to $0\mod\pi$ (an  integer multiple of $\pi$) as  stated by
Levinson's theorem; instead, this difference is equal to $-2^{-3/2}\pi\alpha \,
\mod \, \pi$.
It follows that as the fermion  momentum  $k  \rightarrow  \infty$, the partial
elements of the $S$-matrix tend  to  the  universal limit $\exp(-i 2^{-1/2} \pi
\alpha)$, which is different from unity in the general case.
Hence, the fermion-soliton interaction does not tend to zero from the viewpoint
of unitarity in the ultrarelativistic limit $k \rightarrow \infty$.

The  $\mathbb{CP}^{1}$  Q-lumps  discussed   in   this   present  paper  can be
generalised to a whole  class of K\"{a}hler sigma  models with potential terms,
provided the target manifold has  a  Killing  vector  field  with  at least one
fixed point \cite{abraham_plb_1992}.
In particular, such a generalisation can be done for $\mathbb{CP}^{N-1}$ models
with $N \ge 3$.
The results  obtained  here  for  the $\mathbb{CP}^{1}$  model  can  be  easily
extended to this general case.

\appendix

\section{Partial phase shifts in the semiclassical approximation}     \label{A}

Besides the Born approximation, we can also study the fermion scattering within
the semiclassical approximation \cite{Landau III}.
This approximation is applicable  when the fermion momentum $k \gg 1$, where we
use the dimensionless variables from Sec.~\ref{sec:V}.
In  this  case,  we can  use   normal  form (\ref{III:19g}) of the second-order
differential equation  to  obtain  a  semiclassical  expression for the partial
phase shifts as follows:
\begin{eqnarray}
\delta _{mn}\left( k\right)  &\approx &\int\limits_{\rho_{0}}^{\infty
}\left\{ \left[ k^{2}-\left( m-1/2\right) ^{2}\rho ^{-2}-W\left(\rho \right)
\right]^{1/2}\right.                                                \nonumber
  \\
&&-\left. \left[ k^{2}-\left( m-1/2\right) ^{2}\rho ^{-2}\right]
^{1/2}\right\} d\rho,                                               \label{A:1}
\end{eqnarray}
where the lower limit of integration
\begin{equation}
\rho _{0}\approx \left\vert m-1/2\right\vert /k,                    \label{A:2}
\end{equation}
and the semiclassical potential
\begin{eqnarray}
W\left(\rho \right)  &=& -\frac{\alpha^{2}}{\left(
1+\rho ^{4}\right) ^{2}}+\frac{2\alpha \varepsilon }{1+\rho^{4}}   \nonumber
 \\
&&-\frac{2(2m-3)(\varepsilon +M)\rho ^{2}}{\varepsilon - \alpha
+M+(\varepsilon +M)\rho ^{4}}                                       \nonumber
 \\
&&-\frac{12(\varepsilon +M)(\varepsilon -\alpha +M)\rho ^{2}}{\left(
\varepsilon -\alpha +M+(\varepsilon + M)\rho ^{4}\right)^{2}}       \nonumber
 \\
&&+\frac{2-n}{4}R(\rho).                                            \label{A:3}
\end{eqnarray}
In Eq.~(\ref{A:3}),  the  function $R(\rho)$   is  a  ratio  of  two polynomial
functions, $R(\rho) = A(\rho)/B(\rho)$, where
\begin{eqnarray}
A(\rho) &=&4(2m+3)(\varepsilon - \alpha + M)\rho^{2}+4(2m-1)        \nonumber
 \\
&&\times (\varepsilon + M)\rho^{6}                                  \label{A:4}
\end{eqnarray}
and
\begin{equation}
B(\rho)=\left( 1+\rho ^{4}\right) \left( \varepsilon -\alpha + M +
(\varepsilon + M)\rho^{4} \right).                                  \label{A:5}
\end{equation}

Although the integral in Eq.~(\ref{A:1}) cannot  be calculated analytically, in
the limit of large $k$,  we  can  compute the first few terms of its asymptotic
expansion using approximate analytical methods as
\begin{eqnarray}
\delta_{mn}\left(k\right)&\sim &-2^{-3/2}\pi \alpha+2^{-5/2}\pi     \nonumber
  \\
&&\times \left(2mn -3\right)k^{-1}+O\left[ k^{-3/2}\right].        \label{A:6}
\end{eqnarray}
Note that in the theory of  scattering, the  phase shifts $\delta_{mn}$ are not
unique, but are defined only up to an integer multiple of $\pi$.
The  characteristic feature  of  phase  shifts (\ref{A:6}) is that $\underset{k
\rightarrow\infty}{\lim} \delta_{m n} \left( k \right) = -2^{-3/2} \pi \alpha$,
and is not equal to $0 \, \text{mod} \,\pi$ in the general case.
Note, however, that  $\underset{k \rightarrow \infty}{\lim}\delta_{m n} \left(k
\right) = 0 \,\text{mod}\,\pi$ if $\alpha=2^{3/2} p$, where $p \in \mathbb{Z}$.

In Eq.~(\ref{A:6}), the limiting term $-2^{-3/2} \pi \alpha$ is due to the term
$2 \alpha \varepsilon \left(1 + \rho^{4} \right)^{-1} \sim 2 \alpha k \left(1 +
\rho^{4}\right)^{-1}$  in Eq.~(\ref{A:3}), which increases indefinitely with an
increase in $k$.
This term, in  turn,  arises  from   the   square   of  the  time  component of
covariant derivative (\ref{II:3b}) included in the Dirac equation (\ref{II:4b}).
The nonzero (modulo $\pi$) limiting value of $\delta_{m n}\left(k\right)$ tells
us that in the general case, the fermion-soliton interaction does  not  tend to
zero from the  viewpoint  of  unitarity,  even  in  the ultrarelativistic limit
$k \rightarrow \infty$.

A similar  situation  is   seen   for   the   Coulomb   scattering of fermions.
In this three-dimensional case,  a  term  proportional  to  the  fermion energy
$\varepsilon$ also arises when the Dirac equation is  squared,  and the partial
phase shifts (determined  with account of the logarithmically divergent Coulomb
phase) tend  to nonzero values as $k \rightarrow \infty$.

Finally, we note that  the  Born phase shifts in Eq.~(\ref{IV:11c}) tend to the
same limiting value of $-2^{-3/2}\pi\alpha$  as the  semiclassical phase shifts
in Eq.~(\ref{A:6}).
Furthermore,  the  next-to-leading  order  terms   of   Eqs.~(\ref{IV:11c}) and
(\ref{A:6}) practically coincide for sufficiently large values of $\left\vert m
\right \vert$, when the  accuracy  of the semiclassical approximation improves.

\begin{acknowledgements}

This work was supported by the Russian Science Foundation, grant No 23-11-00002.

\end{acknowledgements}

\end{document}